\documentclass[a4paper,11pt]{article}
\usepackage{jheppub} 
\usepackage[T1]{fontenc} 
\RequirePackage[displaymath]{lineno} 

\usepackage{epsfig}
\usepackage{graphicx}
\usepackage{dcolumn}
\usepackage{bm}
\usepackage{ltablex,booktabs}
\usepackage{overpic}
\usepackage{subfigure}
\usepackage{float}
\usepackage{color}
\usepackage{amsmath}
\usepackage{mathcomp}
\usepackage{mathrsfs}
\usepackage{multirow}
\usepackage{rotating}
\usepackage{amssymb}
\usepackage{gensymb}
\usepackage{amsmath}
\usepackage{tabularx}
\usepackage{makecell}
\usepackage{relsize}\usepackage{tabularx}
\usepackage{booktabs} 
\usepackage{xcolor} 
\def\babar{\mbox{\slshape B\kern-0.1em{\smaller A}\kern-0.1em B\kern-0.1em{\smaller A\kern-0.2em R}}}

\title{Study of \boldmath$\phi\to K\bar{K}$ in the amplitude analysis of \boldmath $D^{+}
  \to K_{S}^{0}K_{L}^{0}\pi^{+}$}

\collaborationImg{\includegraphics[width=0.15\textwidth, angle=90]{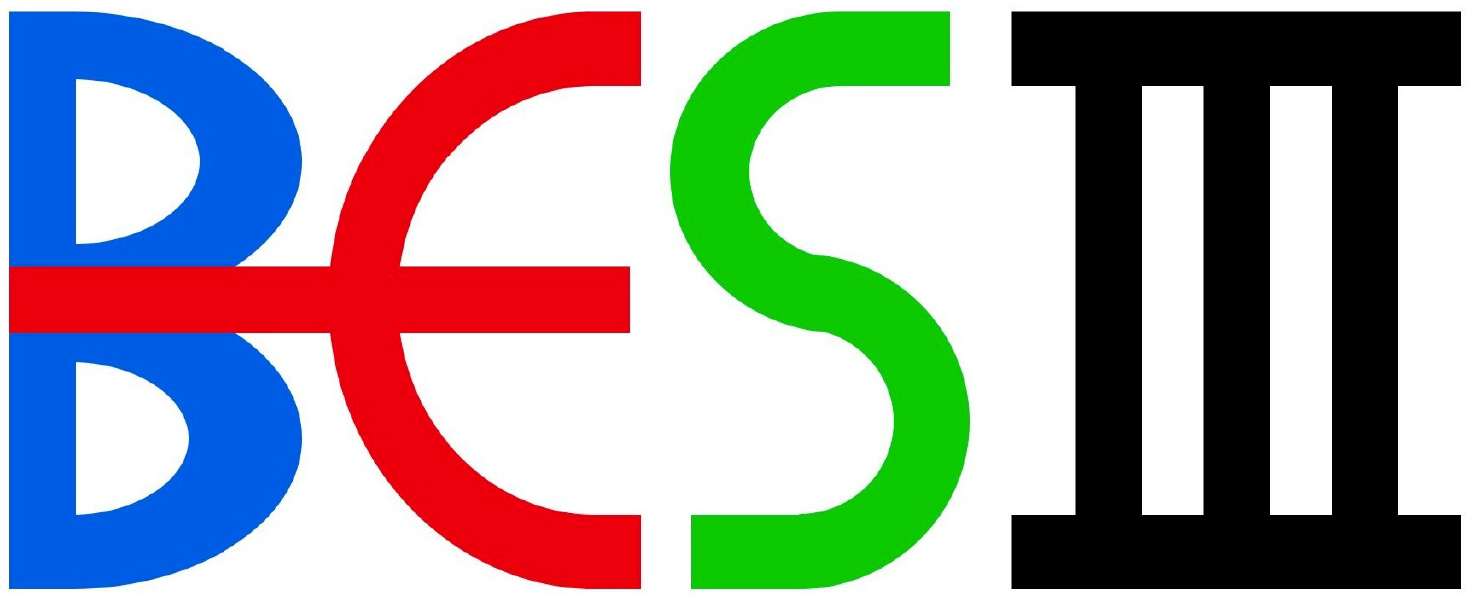}}
\collaboration{BESIII Collaboration}

\date{\today}
\abstract{
We present the first amplitude analysis and branching fraction measurement of $D^{+} \rightarrow K_{S}^{0}K_{L}^{0}\pi^{+}$ decay. The analysis uses a dataset corresponding to an integrated luminosity of 20.3~$\rm fb^{-1}$, which was recorded at a center-of-mass energy 3.773~GeV by the BESIII detector. The measured branching fraction is  
$\mathcal{B}(D^{+} \rightarrow K_{S}^{0}K_{L}^{0}\pi^{+})=(5.758\pm0.085\pm 0.052)\times10^{-3}$, where the first uncertainty is statistical and the second is systematic.
Using the known value of 
${\cal B}(D^+ \to \phi \pi^+,\,\phi \to K^+K^-)$, we determine the relative branching fraction between 
 $\phi \to K_{S}^0K_{L}^0$ and $\phi \to K^+K^-$ to be 
 $\mathcal{B}(D^{+} \to \phi \pi^{+}, \phi \to K_{S}^0K_{L}^0)/\mathcal{B}(D^{+} \to \phi \pi^{+}, \phi \to K^+K^-)= 0.628\pm0.022\pm 0.015\pm0.017$, where the third uncertainty is related to $\mathcal{B}(D^{+} \to \phi \pi^{+}, \phi \to K^+K^-)$.
 This result is significantly lower than the previous world average and is consistent with the isospin expectation for the $\phi$ meson's coupling to charged and neutral kaon pairs. 
}

\arxivnumber{}
\begin{document}
\maketitle
\flushbottom

\section{Introduction}
Measurements of the ratio of neutral-to-charged kaon-pair production rates, from the same initial state, test isospin symmetry. However, the $\phi$ meson has presented a long-standing puzzle: the measured ratio $R^{\phi}_{KK}\equiv\mathcal{B}(\phi \to K_S^0K_L^0)/\mathcal{B}(\phi \to K^+K^-)$ indicates significant isospin-symmetry breaking \cite{PDG}. This anomaly has persisted for decades and lacks a theoretical explanation~\cite{cremmer, BRAMON, Flores, Fischbach, Benayoun}.
Under isospin symmetry $R^{\phi}_{KK}$ is naively unity, but this prediction must be corrected for the difference between the $K^+$ and $K^0$ masses. Nevertheless,  even after considering additional isospin-breaking sources, such as radiative corrections to the 
 $\phi \to K^+K^-$ and $\phi \to K^0\bar{K}^0$ coupling ratio~($g^+/g^0$), theoretical predictions are inconsistent with measurements~\cite{BRAMON, Flores}.  

The ratio $R^{\phi}_{KK}$ has been measured in various experiments using 
different $\phi$-production mechanisms~\cite{PDG}. The most extensive
measurements originate from $e^+e^-$ 
annihilation~\cite{OLYA, HBC72, HBC77, HBC78, CMD2, CMD3, Parrour, Mattiuzzi, Dolinsky}.
In this approach, $R^{\phi}_{KK}$ is extracted from fits to the measured 
$e^+e^-\to K\bar{K}$ cross sections at center-of-mass (c.m) energies near the $\phi$ resonance. These fits must account for interfering amplitudes from other resonances, such as 
$e^+e^-\to \omega,\,\rho \to K\bar{K}$ and non-resonant contributions. 
The extraction of $R^{\phi}_{KK}$ is also inherently sensitive to the $\phi\to e^+e^-$ partial width.
Consequently, these approaches are often limited by uncertainties related to
the cross-section parameterization and the treatment of complex backgrounds. The
reliability of the current experimental average for $R^{\phi}_{KK}$ is further 
challenged by the results reported in Ref.~\cite{Dubn}, which employed an amplitude model consistent with unitarity and analyticity. The reported value of $R_{KK}^{\phi}$ is significantly lower than the experimental world average~\cite{PDG}.
Therefore, alternative measurements of $R^{\phi}_{KK}$ are needed. 

The $\phi$ meson is a common final-state particle in charm-meson decays. Therefore, these decays provide an ideal environment for studying $\phi$-meson properties. Our previous measurements of $\phi$ mesons from $D_s^+$ decays~\cite{lihui, BESIII:2025nou, xiaoyu} observed deviations in both $R^{\phi}_{KK}$ and $R^{\phi}_{\pi\pi\pi} \equiv \mathcal{B}(\phi \to \pi^+\pi^-\pi^0) / \mathcal{B}(\phi \to K^+K^-)$ from the world-average values \cite{PDG}. Using a 20.3~fb$^{-1}$ dataset of electron-positron collisions at a center-of-mass energy $\sqrt{s}=3.773$~GeV~\cite{lum2}, we extend these investigations to $D^+$ decays. In this paper, we perform an amplitude analysis of $D^+ \to K_S^0 K_L^0 \pi^+$ to extract the BF for $D^+ \to \phi \pi^+,~\phi \to K_S^0 K_L^0$. Combined with the known value of ${\cal B}(D^+ \to \phi \pi^+,~\phi \to K^+K^-)$~\cite{PDG}, this measurement provides an estimate of $R^{\phi}_{KK}$. Charge conjugation is implied throughout this paper.

In addition to the $R_{KK}^{\phi}$ measurement, the amplitude analysis of $D^+ \to K_S^0 K_L^0 \pi^+$ probes $D^+ \to \bar{K}^0_{S,L} K^*(892)^+,\,K^*(892)^+ \to K^0_{L,S} \pi^+$ decays, which are singly Cabibbo-suppressed $D \to VP$ decays, where $V$ and $P$ denote vector and pseudoscalar mesons, respectively~\cite{kskpi0,xxh}. Measurements of the branching fraction~(BF) for this decay provide  constraints on dynamical models of charmed meson decays~\cite{Wang, cheng, Qin:2013tje}. 

\section{Detector and data}
\label{sec:detector_dataset}
The BESIII  detector~\cite{Ablikim:2009aa} records symmetric $e^+e^-$ collisions provided by the BEPCII storage ring~\cite{Yu:IPAC2016-TUYA01} in the center-of-mass energy range from 1.84 to 4.95~GeV, with a peak luminosity of $1.1 \times 10^{33}\;\text{cm}^{-2}\text{s}^{-1}$ achieved at $\sqrt{s} = 3.773\;\text{GeV}$. BESIII has collected large data samples in this energy region~\cite{Ablikim:2019hff, EcmsMea, EventFilter}. The cylindrical core of the BESIII detector covers 93\% of the full solid angle and consists of a helium-based  multilayer drift chamber~(MDC), a plastic scintillator time-of-flight system~(TOF), and a CsI(Tl) electromagnetic calorimeter~(EMC), which are all enclosed in a superconducting solenoidal magnet providing a 1.0~T magnetic field. The solenoid is supported by an octagonal flux-return yoke with resistive plate counter muon identification modules interleaved with steel. The charged-particle momentum resolution at 1 GeV$/c$ is $0.5\%$, and the ionization energy loss (${\rm d}E/{\rm d}x$) resolution in the MDC is $6\%$ for electrons from Bhabha scattering. The EMC measures photon energies with a resolution of $2.5\%$ ($5\%$) at $1$~GeV in the barrel (end-cap) region. The time resolution in the TOF barrel region is 68~ps, while that in the end-cap region was 110~ps.  The end-cap TOF system was upgraded in 2015 using multigap resistive plate chamber technology, providing a time resolution of 60~ps, which benefits 86\% of the data used in this analysis~\cite{etof1,etof2,etof3}.

The data sample with a total integrated luminosity of $20.3 \, \mathrm{fb}^{-1}$, collected at the center-of-mass energy of $\sqrt{s} = 3.773\, \mathrm{GeV}$, is used in this analysis. The $\psi(3770)$ predominantly decays to $D^+D^-$ or $\bar{D}^0 D^0$ pairs without additional hadronic activity, providing an ideal environment for studying $D$ meson decays with the double-tag (DT) technique~\cite{DTmethod}.
In this method, a single-tag (ST) candidate requires the reconstruction of a single ${D}^{-}$ meson through one of the three hadronic decay modes: $D^-\to K^{+}\pi^{-}\pi^{-}$, $K^{0}_{S}\pi^{-}$, and $K_{S}^{0}\pi^{+}\pi^{-}\pi^{-}$. These decays are termed tag modes. In the tag-mode selection, no requirements are made on reconstructed tracks and electromagnetic-calorimeter showers not used in the selection. In the DT sample, the $D^{+}$ meson is reconstructed through the signal-decay mode $K^0_{S}K^0_{L}\pi^{+}$ and the associated ${D}^{-}$ through one of the tag modes.

Monte Carlo (MC) simulated data samples, produced with a {\sc
geant4}-based~\cite{geant4} software package, 
are used to determine detection efficiencies and estimate backgrounds. 
The simulation includes the geometric description of the BESIII detector, the
detector response, and models the beam-energy spread and initial-state radiation (ISR) in the $e^+e^-$
annihilations with the generator {\sc kkmc}~\cite{KKMC1,KKMC2}. The inclusive MC sample includes the production of $D\bar{D}$ pairs (with quantum-coherence effects for the neutral $D^0 \bar D^0$ channels), the non-$D\bar{D}$ decays of the $\psi(3770)$, the ISR production of the $J/\psi$ and $\psi(3686)$ states, and the continuum processes incorporated in {\sc kkmc}~\cite{KKMC1,KKMC2}.
All particle decays are modeled with {\sc
evtgen}~\cite{EVTGEN1,EVTGEN2} using BFs 
either taken from the
Particle Data Group (PDG)~\cite{PDG}, when available,
or otherwise estimated with {\sc lundcharm}~\cite{LUNDCHARM1,LUNDCHARM2}.
Final-state radiation
from charged final-state particles is incorporated using {\sc
photos}~\cite{PHOTOS}.
A phase-space (PHSP) MC sample is generated with a uniform distribution for the decay $D^+ \to  K_S^0 K_L^0 \pi^+$ , which is used to determine the detection efficiency function and calculate the normalization integral used in the determination of the amplitude-model parameters in the fit to data.
A signal MC sample, generated according to the results of the amplitude analysis for the decay $D^+ \to  K_S^0 K_L^0 \pi^+$ , is used to check the fit performance, calculate the goodness of fit and estimate the average DT efficiency in the BF measurement.

\section{Event selection}
\label{ST-selection}

We use identical event selection for the amplitude analysis and branching fraction measurement.
The $D^-$ candidates are constructed from individual $\pi^\pm, K^\pm$ and $K_S^0$ candidates with the following selection criteria. 
 
Charged tracks detected in the MDC are required to be within a polar angle ($\theta$) range of $|\rm{cos\theta}|<0.93$, where $\theta$ is defined with respect to the $z$-axis, which is the symmetry axis of the MDC. For charged tracks, the distance of closest approach to the interaction point (IP) must be less than 10\,cm along the $z$-axis,  and less than 1\,cm in the transverse plane.

Particle identification~(PID) for charged tracks combines measurements of the ionization energy loss d$E$/d$x$ and the flight time in the TOF to form likelihoods $\mathcal{L}(h)~(h=K,\pi)$ for each hadron $h$ hypothesis.
Charged kaons and pions are identified by comparing the likelihoods, $\mathcal{L}(K)>\mathcal{L}(\pi)$ and $\mathcal{L}(\pi)>\mathcal{L}(K)$, respectively.

The $K_{S}^0$ candidates are selected from all possible pairs of tracks with opposite charges and the distances of the charged tracks to the interaction point along the beam direction are required to be within 20 cm. The selected charged tracks are assigned as pions and no further PID requirements are applied.
A primary vertex and a secondary vertex are reconstructed, and the decay length between two vertices is required to be greater than twice its uncertainty. The $\chi^2$ of the vertex fit must be less than 100 and the invariant mass of the $\pi^{+}\pi^{-}$ pair ($M_{\pi^{+}\pi^{-}}$) is required to be in the range $[0.487, 0.511]$ GeV/$c^2$ to form the candidate $K_{S}^0$ particles.



The tag-mode $D^-$ mesons are selected using two kinematic variables, the beam-constrained mass $M_{\rm BC}$ and the energy difference $\Delta E$:
\begin{equation}
\begin{aligned}
	&M_{\rm{BC}} = \sqrt{E_{\rm{beam}}^2 /c^4-|\vec{p}_{D}|^2 /c^2}, \\
	&\Delta E = E_{D} - E_{\rm{beam}},
\end{aligned}
\label{MBC_DeltaE}
\end{equation}
where $E_{\rm beam}$ is the  beam energy. Here, $\vec{\mathbf{p}}_{D^-}$ and $E_{D^-}$ are 
the reconstructed momentum and energy of the $D^-$ candidate, respectively. Correctly-reconstructed 
$D^-$ mesons appear as a peak at the known $D^-$ mass~\cite{PDG} in the 
$M_{\rm BC}$ distribution and at zero in the $\Delta{E}$ distribution. If we obtain multiple $D^-$ candidates in an event, the candidate with the minimum 
$\lvert \Delta E \rvert$ is retained. All $D^-$ candidates must satisfy 
$1.863 < M_{\rm BC} < 1.877$~GeV/$c^{2}$.  The criteria selecting final-state particles, as well as 
$\Delta E$ requirements, are identical to those reported in Refs.~\cite{BESIII:2024ncc, ref:kspieta}.

We identify the signal $D^{+} \to K^0_{S}K^0_{L}\pi^{+}$ from the unused tracks and clusters which remain after reconstructing the tag mode. 
Exactly three tracks are required, two positively-charged and one negatively-charged. At least one pair of oppositely-charged particles must be consistent with
$K_S^0\to \pi^+\pi^-$ decay; the other charged particle must be identified as a pion. The track, $K^0_S$, and PID selection criteria are the same as for the ST side. A kinematic fit improves the four-momentum resolution of the final-state particles. The fit constrains the total final-state four-momentum to the initial-state $e^+e^-$ four-momentum, as well as the invariant 
masses of $D^{\pm}$ and $K_S^0$ candidates to their known masses~\cite{PDG}, 
while the missing four-momentum is considered as a $K_L^0$ candidate.  The fit must converge and satisfy $\chi^2<100$. If both combinations satisfy 
these criteria, the one with the lowest $\chi^2$ value is selected. The missing-mass is 
\begin{eqnarray}
\begin{aligned}
  M_{\rm miss}=\sqrt{\frac{1}{c^2}(p_{e^+e^-}-p_{\rm tag}-p_{K_S^0}-p_{\pi^+})^2}\,, 
  \label{mm2}
\end{aligned}
\end{eqnarray}
where $p_{e^+e^-}$, $p_{\rm tag}$, and $p_{i}~(i=K_{S}^0,~\pi^+)$ are  the four-momenta of the initial-state $e^+e^-$, the tag mode, and the final-state signal particles, respectively. 
For correctly reconstructed DT candidates, $M_{\rm miss}$ will peak at the known $K_{L}^0$ mass~\cite{PDG}.

The decay $D^+\to K_S^0\pi^+\eta$ and $D^{+} \to K_{S}^{0}K_{S}^{0}\pi^{+}$, with one 
$K_{S}^{0} \to \pi^{0}\pi^{0}$, form peaking backgrounds in $M_{\rm miss}$ at the known
$\eta$ and $K_L^0$ masses~\cite{PDG}, respectively.
To suppress these backgrounds, we veto events in which either a 
$\pi^0$ or $\eta$ candidate is reconstructed in the $\gamma\gamma$ final state. The $\pi^0$ and $\eta$ candidate selection criteria are identical to those used in Refs.~\cite{BESIII:2024ncc, ref:kspieta}. An additional source of peaking background is from $D^{+} \to K_{L}^{0}\pi^{+}\pi^{+}\pi^{-}$ decays, in which a pair of oppositely charged tracks form a fake $K_{S}^0$ candidate.

Candidates with $0.469<M_{\rm miss}<0.515$~GeV/$c^2$ are selected for the amplitude analysis. The resulting sample contains $11387$ events with a purity 
$f_s = (67.5\pm 0.5)$\%, which is determined by fitting the 
$M_{\rm miss}$ distribution, as shown in Fig.~\ref{fig:DT_fit}. 
The background contribution is separated into three components:
(1) the peaking background from $D^+\to K_L^0\pi^+\pi^+\pi^-$ and $D^+\to K_S^0K_S^0\pi^+$ decays; (2) $D^+\to K_S^0\pi^+\eta$ decays; and (3) combinatorial 
background. The PDFs of signal, peaking background, and 
$D^+\to K_S^0\pi^+\eta$ components are modeled by MC-simulated shapes. These simulations are based on their respective 
amplitude analyses~\cite{BESIII:2024ncc, ref:kspieta, BESIII:2019ymv}.
The shape of the combinatorial background is simulated by inclusive MC. 
All shapes, except the combinatorial background, are convolved with a Gaussian function to account for any data-MC resolution difference; the width is fixed to that measured from a control sample of $D^+\to K_S^0\mu^+\nu_\mu$ decays. 
The yields of signal, $D^+\to K_S^0\pi^+\eta$, 
and the combinatorial background are free parameters in the fit, while
the peaking-background yield is Gaussian-constrained to the expected yield $119\pm 4$ estimated from the known BFs~\cite{BESIII:2024ncc, BESIII:2019ymv}.

\section{Amplitude analysis}
\label{Amplitude-Analysis}
\label{sec:pwa-select}


The amplitude analysis uses an isobar model~\cite{covariant-tensors}.
The total amplitude of the signal $\mathcal{M}=\begin{matrix}\sum \rho_{n}e^{i\varphi_{n}}A_{n}\end{matrix}$, is described by a coherent sum over the decay amplitudes $A_{n}$ of $n$ intermediate processes, with relative amplitudes $\rho_{n}$ and phases $\varphi_{n}$. The decay amplitude of the $n^{\rm th}$ intermediate state $A_{n}$ depends on the four-momenta of the final-state particles and the spin of the intermediate resonance: $A_{n} = P_{n}S_{n}F_{n}^{r}F^{D}$, where $P_{n}$ is the Breit-Wigner propagator~\cite{RBW}, $S_{n}$ is the spin factor formulated with covariant tensors~\cite{covariant-tensors}, and $F_{n}^{r}~\left(F^D\right)$ is the
Blatt-Weisskopf barrier factor of the resonance  ($D^{+}$ meson)~\cite{Blatt}.

A maximum-likelihood fit determines the parameters of the amplitude model. We construct a combined probability density function~(PDF) for the signal and background hypotheses that depends on the four-momenta of the final-state particles.  
The signal PDF is based on the total 
amplitude $\mathcal{M}$. The efficiency-corrected background PDF $B_\epsilon$ is
estimated from a background shape
derived from the inclusive MC samples using the XGBoost package~\cite{xgboost1,xgboost2}, following the method described in Ref.~\cite{BESIII:2025sea}.
This background PDF is added to the
signal PDF incoherently. The log-likelihood function $\ln\mathcal{L}$ is written as
\begin{eqnarray}
  \begin{aligned}
       \sum_{j} \ln\left[\frac{f_s\left|\mathcal{M}(p)\right|^{2}}{\int \epsilon(p)\left|\mathcal{M}(p)\right|^{2}\,R_{3}(p)dp}
          +\frac{(1-f_s)B_\epsilon(p)}{\int  B_\epsilon(p)\,R_{3}(p)dp}\right], 
  \end{aligned}
\end{eqnarray}
where $j$ runs over the selected events and $R_{3}$ 
is the three-body phase space, and a kinematic fit is performed with an additional $K_L^0$ mass constraint added to the original constraints, with the final-state particle four-momenta $p$ used for the amplitude analysis.
The detection efficiency $\epsilon$ is 
determined with an MC sample of $D^+\to K_S^0K_L^0\pi^+$ uniformly distributed 
over the Dalitz plot. 
The normalization integral in the denominator accounts for detector resolution using an MC technique as well as efficiency differences between data and MC related to tracking, PID, and $K_S^0$ reconstruction 
as described in Refs.~\cite{ref:Kspipi0, ref:KsKpipi, ref:KsKpi0}.

\begin{figure}[!t]
  \centering
  \includegraphics[width=0.46\textwidth]{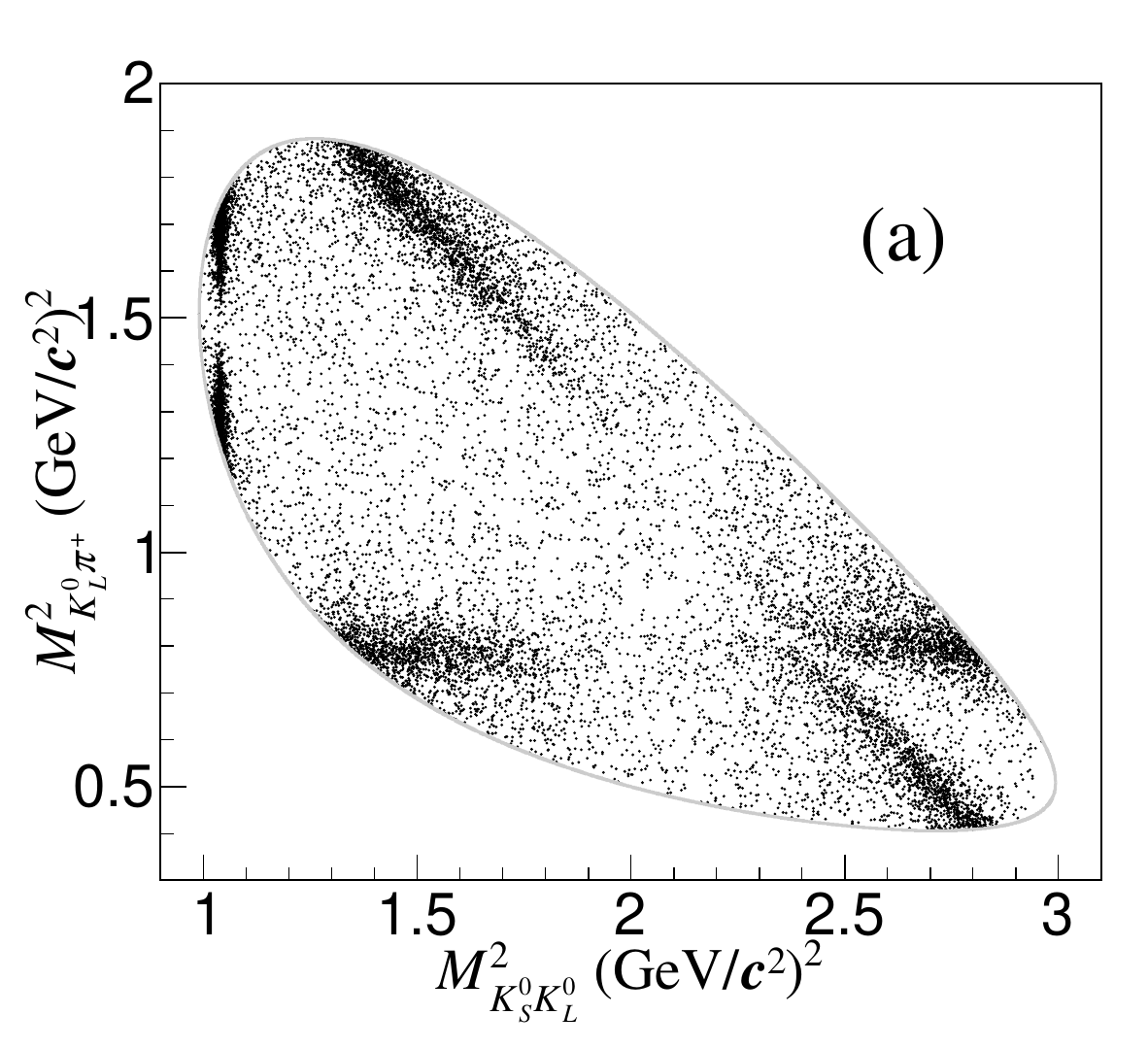}
    \includegraphics[width=0.46\textwidth]{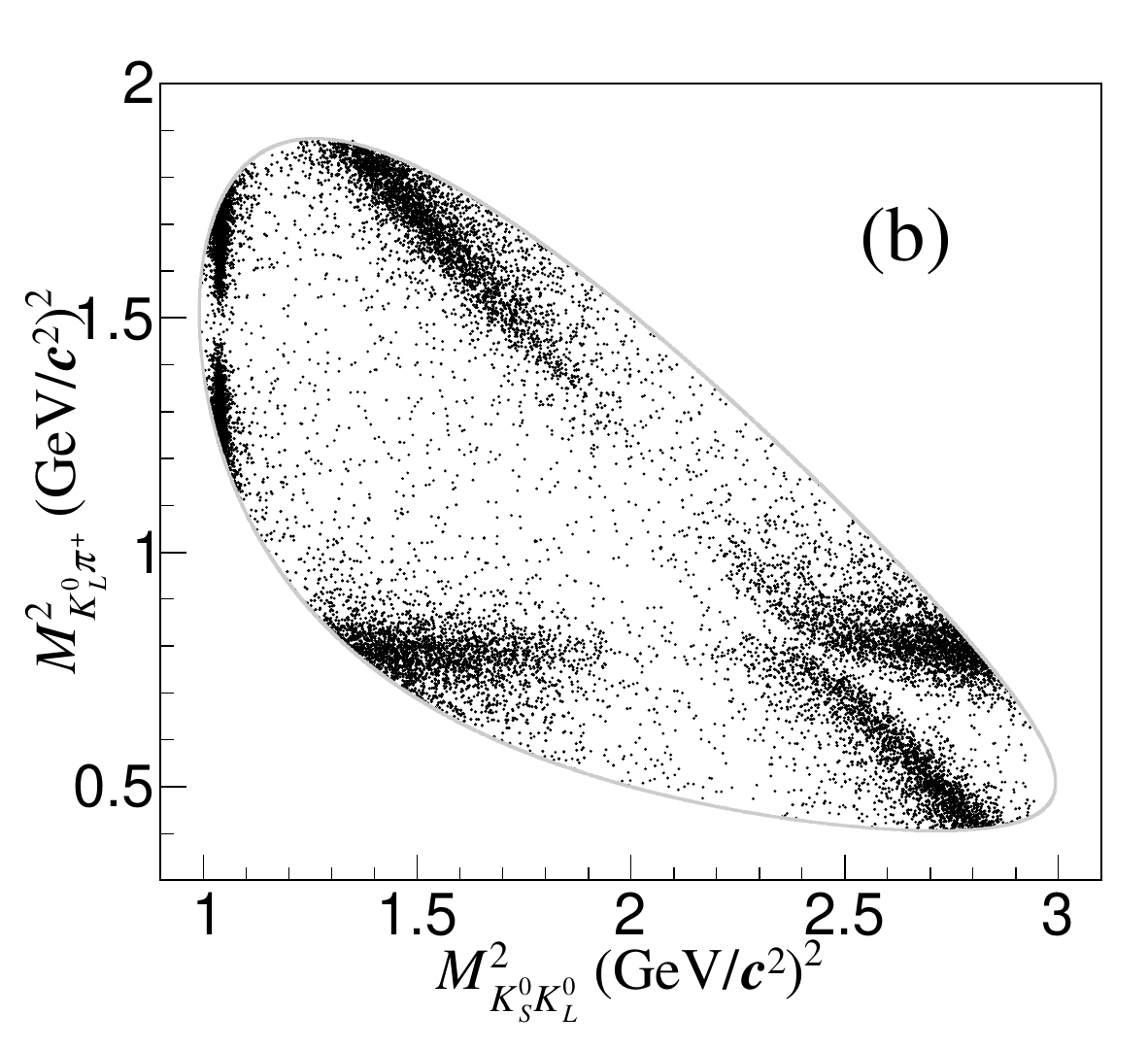}
    \caption{Dalitz plot of $M^{2}_{K_S^0K_L^0}$ versus $M^{2}_{K_L^0\pi^+}$ for (a) the data sample and (b) the simulated samples with signal events generated by the results of amplitude analysis and the background events from the inclusive MC.  
    The gray curve indicates the  kinematic boundary.} 
    \label{fig:dalitz}
\end{figure}

 The Dalitz plot of $M^2_{K_S^0K_L^0}$ versus $M^2_{K_L^0\pi^+}$ from data is shown in
Fig.~\ref{fig:dalitz}(a). The vertical band on the left is caused by the process $D^+\to \phi \pi^+$, the horizontal and diagonal bands around
0.8~GeV$^2$/$c^4$ are $D^+\to K_L^0K^{*}(892)^+$ and $D^+\to K_S^0K^{*}(892)^+$, respectively.

\begin{figure*}[!t]
  \centering
  \includegraphics[width=0.48\textwidth]{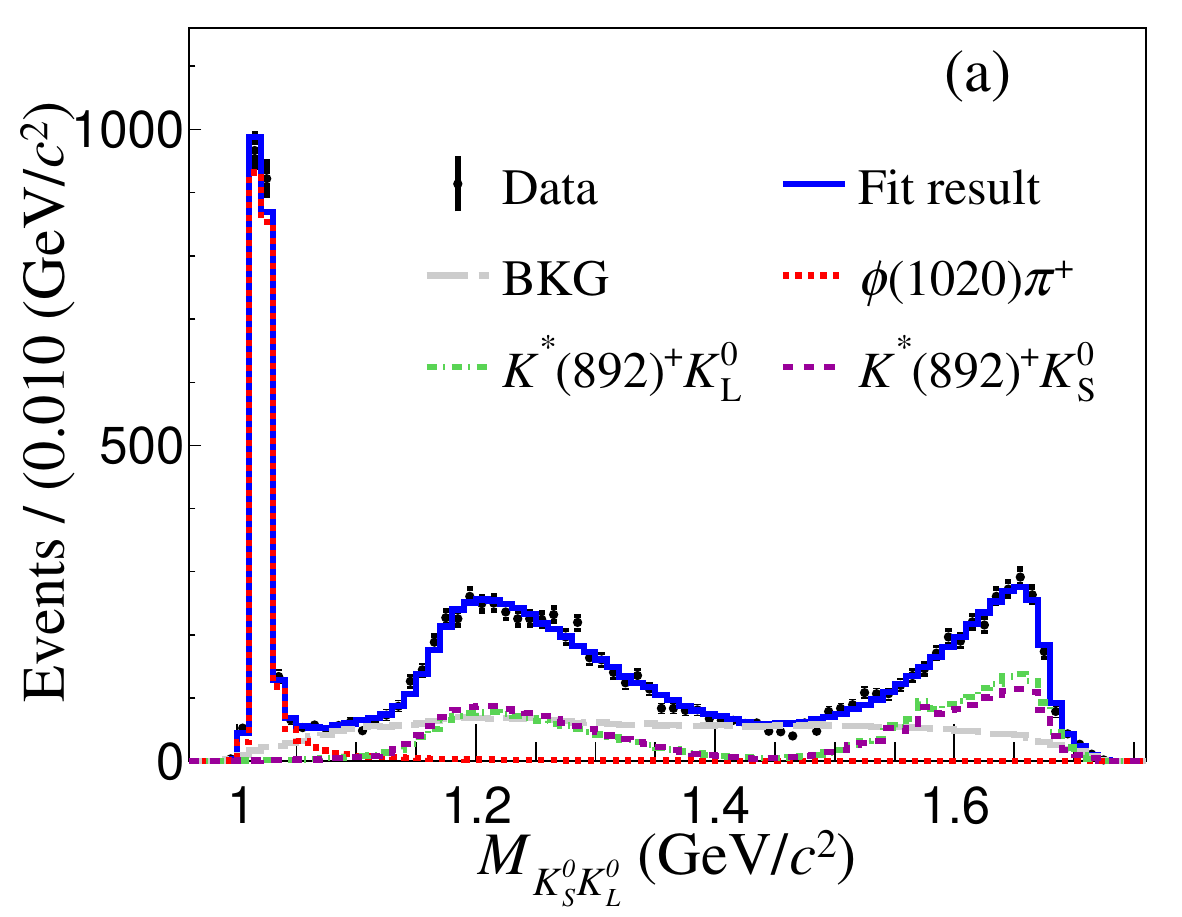}
  \includegraphics[width=0.48\textwidth]{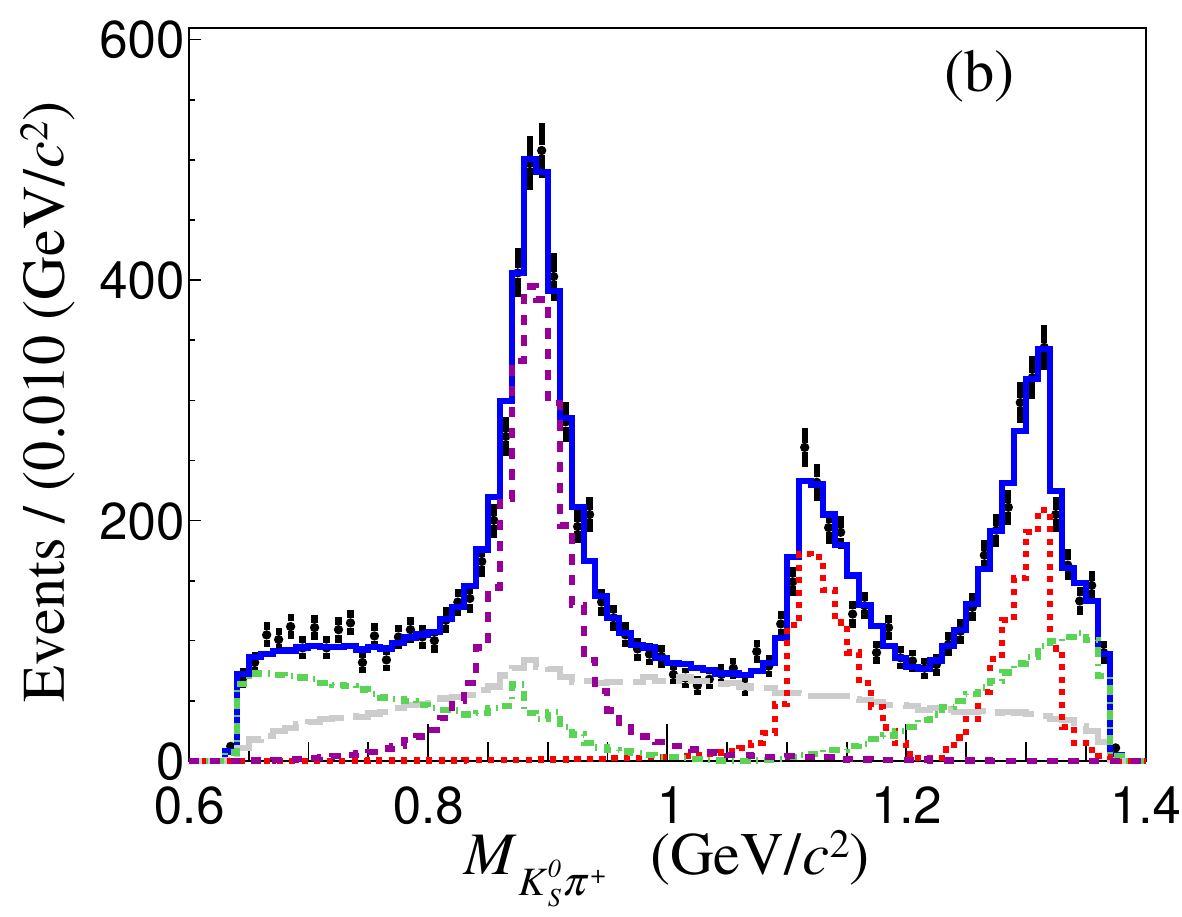}\\
  \includegraphics[width=0.48\textwidth]{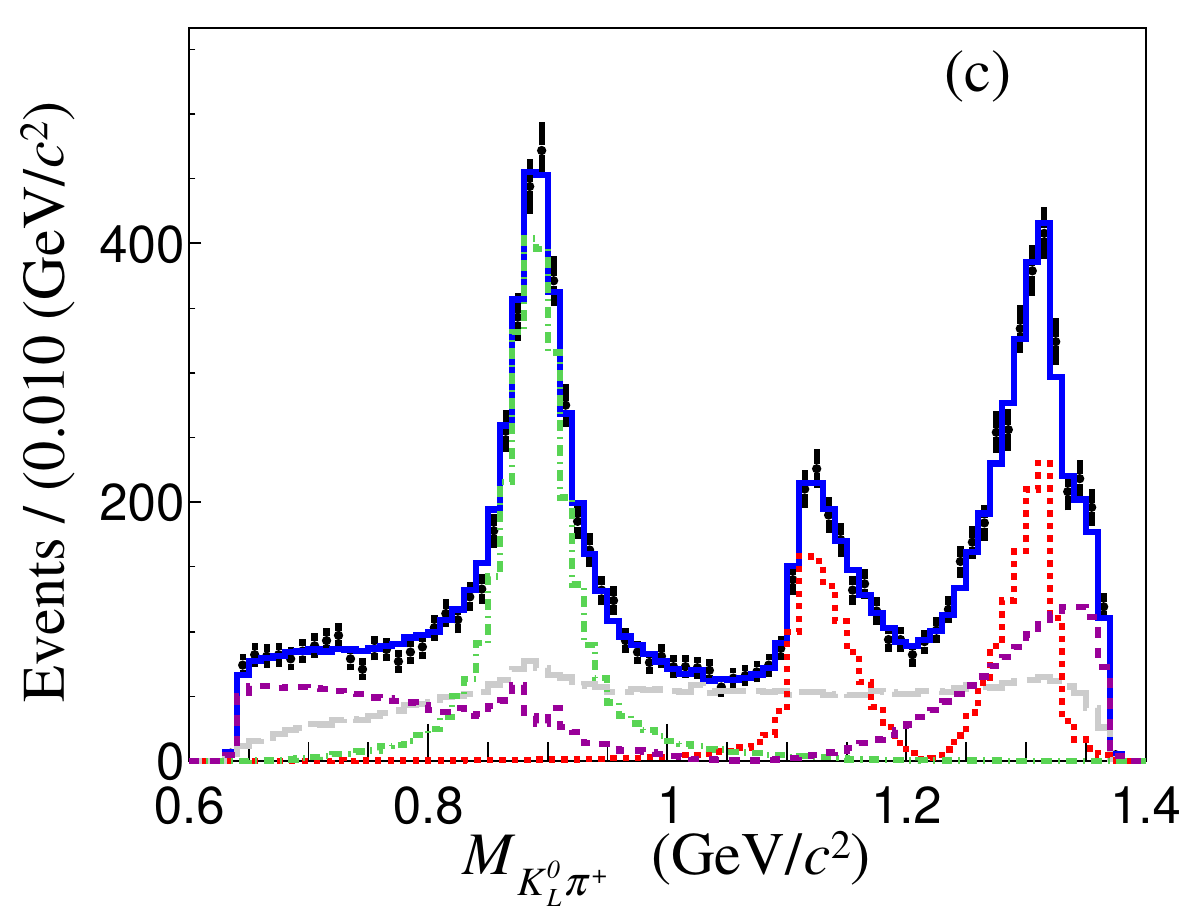}
  \caption{ Distributions of (a) $M_{K_S^0K_L^0}$, (b) $M_{K_S^0\pi^+}$ and (c) $M_{K_L^0\pi^{+}}$ from the nominal fit. 
  The data samples are represented by the points with error bars, the fit results by the blue lines, and the backgrounds by the gray dashed lines. Colored dashed lines show the individual components of the fit model. 
    }
\label{dalitz-projection}
\end{figure*}

The $D^+\to \phi \pi^+$ amplitude is taken as the reference, with fixed values $\rho_{\phi\pi}=1$ and $\varphi_{\phi\pi}=0$. The $\rho_n$ and $\varphi_n$ values for  other resonances are determined relative to it. The purity value $f_s$ is fixed to the central value determined from the fit to $M_{\rm miss}$.
Other possible contributing resonances such as $K^{*}(1410)^{+}$,
$K^{*}_{0}(1430)^{+}$, $K^{*}_{2}(1430)^{+}$, $(K_S^0\pi^{+})_{\mathcal{S}-\rm wave}$~\cite{KpiLASS, Kmatrix}, $(K_L^0\pi^{+})_{\mathcal{S}-\rm wave}$, $\phi(1680)$, $\rho(1450)$, and $\rho(1700)$ are added to the
fit one at a time.
The masses and widths of all resonances are fixed to their known values~\cite{PDG}.
The statistical significance of each new amplitude
is calculated from the log-likelihood change accounting for the differing number of 
degrees of freedom. Various combinations of these resonances are tested. 
The only significant contributions found are from $\phi \pi^+$, $K_L^0K^{*}(892)^+$, and $K_S^0K^{*}(892)^+$; no other contribution have a significance greater than $3\sigma$.
The Dalitz plot of the signal MC sample
generated based on the result of the amplitude analysis is shown in Fig.~\ref{fig:dalitz}(b).
The mass projections of the fit are shown in ~Fig.~\ref{dalitz-projection}.

The contribution of the $n^{\rm th}$ amplitude relative to the total BF is quantified
by the fit fraction~(FF) defined as
${\rm FF}_{n} = \int \left|\rho_{n}A_{n}\right|^{2}R_3 dp/\int\left|\mathcal{M}\right|^{2}R_3dp$.
The FFs of amplitudes and the phase differences relative to the reference process are listed in Table~\ref{fit-result}, where the significances for each intermediate process all exceed $10\sigma$. The phase difference of nearly $\pi$ radians between $D^{+} \to K_{S}^0K^{*}(892)^{+}$ and $D^{+} \to K_{L}^0K^{*}(892)^{+}$ amplitudes leads to destructive
interference between them. As a result of this interference the total fit fraction is $\left(107.7\pm 1.5\right)\%$.


The uncertainties related to input parameters are estimated with the following parameter variations:
the masses and widths of the $\phi$ and $K^{*}(892)^{+}$ varied by their uncertainties~\cite{PDG}; the radii of the Blatt-Weisskopf
barrier factors varied from their nominal values of $F^D=5$~GeV$^{-1}$ and $F^r_n=3$~GeV$^{-1}$ by $\pm 1$~GeV$^{-1}$; and $f_s$ varied within its statistical
uncertainty. We include a systematic uncertainty from the difference between the nominal fit and one with an alternative background PDF derived using the relative fraction of $q\bar{q}$ background processes varied by the uncertainties in the known cross section~\cite{BES:2007cev}.
The uncertainties related to the peaking background model are estimated by varying these components within the uncertainties of their measurements~\cite{BESIII:2024ncc, BESIII:2019ymv}. 
The systematic uncertainties from PID, tracking, and $K_S^0$ reconstruction efficiencies are evaluated based on Ref.~\cite{ke}.
In addition, we estimate the systematic uncertainty related to model selection by taking the difference between the results from the nominal fit and fits with potential intermediate resonances introduced individually. 
By adding these contributions in quadrature we obtain the total systematic uncertainties, which are given in Table~\ref{fit-result}.
\begin{table*}[!t]
  \caption{Phases, FFs, and BFs of intermediate processes in $D^+\to K_S^0K_L^0\pi^+$. 
  The first and second
    uncertainties are statistical and systematic, respectively. 
    }
  \label{fit-result}
  \begin{center}
    \begin{tabular}{lcccc}
      \hline \hline
            Amplitude  &\phantom{0} Phase~(rad) & FF~(\%)  & BF~($10^{-3}$)                \\
      \hline
      $D^{+} \to \phi \pi^{+}$      & 0.0(fixed)                     & \phantom{0} $29.14 \pm 0.97 \pm 0.65 $ &\phantom{0} $1.68\pm 0.06 \pm 0.04 $ &\phantom{0}  \\ 
      $D^{+} \to K_{L}^{0}K^{*}(892)^{+}$      &\phantom{0}$ 2.32 \pm 0.09 \pm 0.09$  &\phantom{0} $40.40 \pm 0.83 \pm 0.26$ &\phantom{0}  $2.33\pm 0.06 \pm 0.03$ &\phantom{0} \\  
      $D^{+} \to K_{S}^{0}K^{*}(892)^{+}$         &\phantom{0}$ 5.48 \pm 0.09 \pm 0.09$  &\phantom{0} $38.03 \pm 0.79\pm 0.25$ &\phantom{0}  $2.19\pm 0.06 \pm 0.03$ &\phantom{0} \\          
      \hline \hline
    \end{tabular}
  \end{center}
\end{table*}

\section{Branching fraction measurement}

\begin{figure}[t!]
  \begin{center}
    \includegraphics[width=0.6\textwidth]{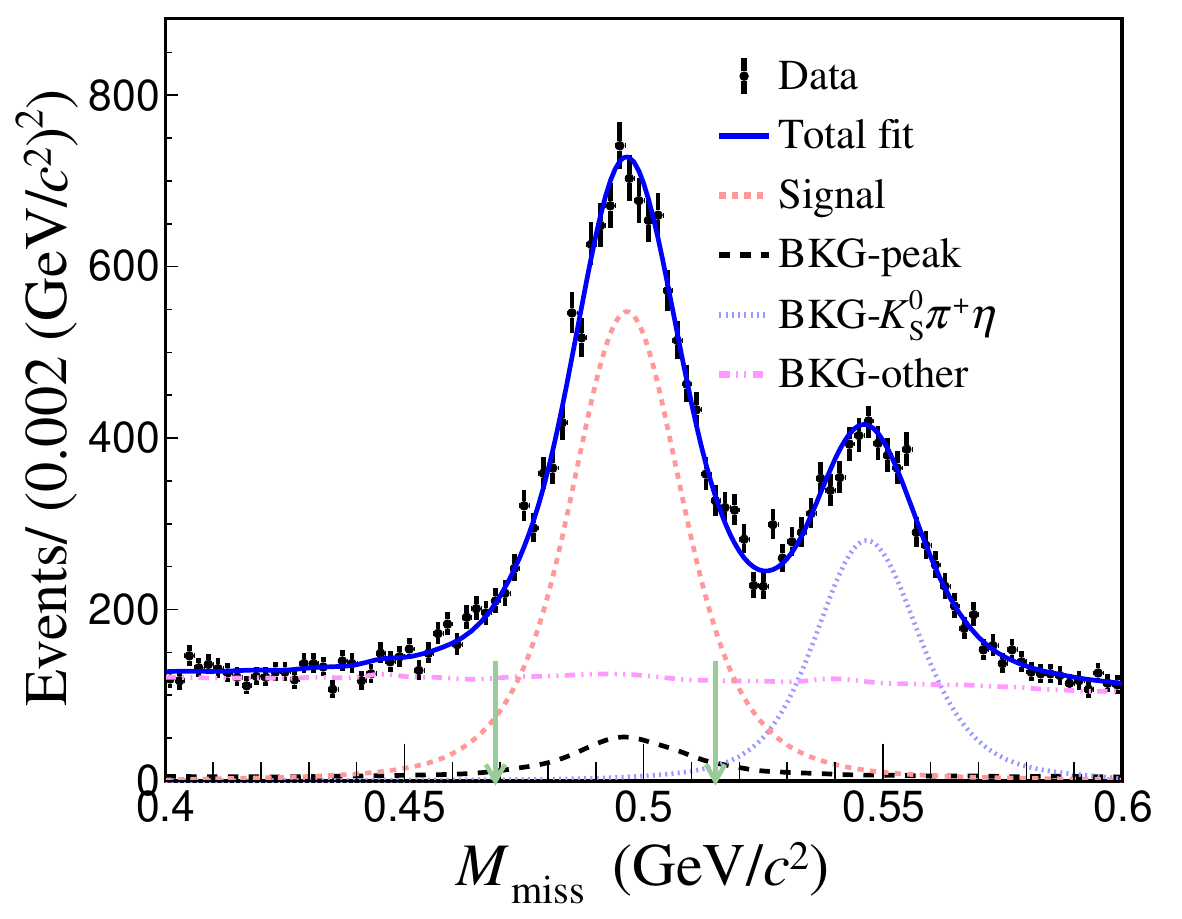}
    \caption{
    Fit to the $M_{\mathrm{miss}}$ distribution. 
Data points with error bars are shown. 
The total fit is represented by the solid blue line and 
the signal by the dotted red line.
Contributions from $D^{+} \to K_{S}^{0}K_{S}^{0}\pi^{+}$ and $D^{+} \to K_{L}^{0}\pi^{+}\pi^{+}\pi^{-}$~(BKG-peak), 
$D^+ \to K_{S}^{0} \pi^+ \eta$~(BKG-$K_{S}^{0} \pi^+ \eta$), and 
other backgrounds~(BKG-other) are displayed as the 
dashed black, long-dashed cyan,  
and dash-dotted purple lines, respectively. 
The signal region for the amplitude analysis is indicated by a pair of green arrows.
     }
    \label{fig:DT_fit}
  \end{center}
\end{figure}

The BF of 
$D^{+} \to K_{S}^{0}K_{L}^{0}\pi^{+}$ is given by~\cite{ref:Kspipi0, ref:KsKpipi}
\begin{eqnarray} \begin{aligned}
\mathcal{B}_{\text{sig}}=\frac{N_{\text{total,sig}}^{\text{DT}}}{\mathcal{B}_{\rm sub}f_{\rm corr}\sum_{   
        \alpha} N_{\alpha}^{\text{ST}}\epsilon^{\text{DT}}_{\text{sig},
        \alpha}/\epsilon_{\alpha}^{\text{ST}}}\,, \label{eq:Bsig-gen}
\end{aligned} \end{eqnarray} 
where $\alpha$ runs over the various tag modes, $\mathcal{B}_{\rm sub}$
represents the BF of $K_S^0 \to \pi^+ \pi^-$, and $f_{\rm corr}=1.002\pm 0.005$ is the
correction factor for data-MC efficiency difference. The ST yields
$N_{\alpha}^{\text{ST}}$ and the DT yield $N_{\text{total,sig}}^{\text{DT}}$
are determined from likelihood fits to the $M_{\rm BC}$ and $M_{\rm miss}$
distributions, respectively. The fit $M_{\rm miss}$ distribution is shown in Fig.~\ref{fig:DT_fit}.
The corresponding efficiencies $\epsilon$ are obtained by analyzing the inclusive MC samples, with the signal decays $D^+\to K_S^0K_L^0\pi^+$ generated according to the nominal amplitude-analysis model reported in this paper. The details of ST yields, ST efficiencies, and DT efficiencies are provided in  Table~\ref{ST-eff}. 
The total ST yields and the DT yields are
$(7,001.8\pm2.9)\times10^3$ and $9,177\pm 134$, respectively. These yields give 
$\mathcal{B}\left(D^+\to K_S^0K_L^0\pi^+\right)=
(5.758\pm0.085\pm0.052)\times10^{-3}$.

\begin{table}[t!]
  \caption{ST yields ($N_{\rm ST}$), ST efficiencies ($\epsilon_{\rm ST}$),
  and DT efficiencies ($\epsilon_{\rm DT}$). The efficiencies include the 
  sub-resonance decays. The uncertainties are statistical.}
    \begin{center}
      \begin{tabular}{lccc}
        \hline   \hline      
        Tag mode & $N_{\rm ST} (10^{3})$        & $\epsilon_{\rm ST}(\%)$   & $\epsilon_{\rm DT}(\%)$ \\
        \hline
$K^{+}\pi^{-}\pi^{-}$ & $5552.8\pm 2.5$ & $51.10\pm0.01$ & $16.67\pm 0.02$ \\
$K_S^0\pi^{-}$        & $657.8 \pm0.9$   & $51.53\pm0.01$ & $18.24\pm0.04 $ \\
$K_S^{0}\pi^+\pi^{-}\pi^{-}$ & $791.2\pm1.1$ & $29.63\pm0.01$ & $9.20\pm0.02 $ \\
        \hline \hline 
      \end{tabular} 
    \end{center}
    \label{ST-eff}
\end{table}

The following systematic uncertainties are considered in the BF measurement. The relative uncertainty in the total number of ST $D^-$ mesons is evaluated 
to be 0.3\% where the dominant source is the parameterization of the fit. The uncertainty from the signal shape is estimated by varying the fixed width by $\pm 1\sigma$ and is found to be negligible. The uncertainty related to the combinatorial background shape is evaluated by repeating the fit 
with the relative fractions of $q\bar{q}$ 
varied by 
$\pm 5\%$, which is the uncertainty in the measured cross sections~\cite{BES:2007cev}.
The $\pi^+$ PID, tracking efficiencies, and the 
$K_{S}^{0}$ reconstruction are studied with DT samples of hadronic $D$ decays. 
The data-MC efficiency ratios for $\pi^+$ PID and tracking are 
determined to be $0.997\pm0.001$ and $0.996\pm0.001$, respectively. That for the $K_{S}^{0}$ reconstruction is $0.995\pm0.002$. 
The uncertainty related to $\pi^0$ and $\eta$-veto efficiency is studied in a control sample of $D^+\to K_S^0\mu^+\nu_\mu$ decays; this gives a data-MC efficiency ratio of $1.014\pm 0.004$.
These individual factors are multiplied to determine $f_{\rm corr}$ and its associated systematic uncertainty of $0.5\%$.
The uncertainty related to the signal-MC model 
is studied by varying the amplitude-model
parameters according to their covariance matrix, which incorporates both statistical and systematic uncertainties. The change in signal
efficiency is 0.2\%.
The relative uncertainty related to the input value of the $K_S^0 \to \pi^+ \pi^-$ BF is 0.1\%~\cite{PDG}.
The total uncertainty, 0.9\%, is determined by adding all contributions in quadrature.
\section{Summary}
We present the first amplitude analysis and BF measurement 
of the hadronic decay $D^{+} \to K_{S}^0K_{L}^0\pi^{+}$ using 20.3~fb$^{-1}$ of 
$e^+e^-$ annihilation data taken at the c.m. energy $\sqrt{s}= 3.773$~GeV. We measure $\mathcal{B}\left(D^{+} \to K_{S}^0K_{L}^0\pi^{+}\right)=(5.758\pm0.085\pm0.052)\times10^{-3}$.
BFs of the intermediate processes are calculated via 
${\cal B}_i={\rm FF}_i\times{\cal B}(D^{+} \to K_{S}^0K_{L}^0\pi^{+})$,  which are given in Table~\ref{fit-result}. 

\begin{figure}[t!]
  \begin{center}
    \includegraphics[width=0.6\textwidth]{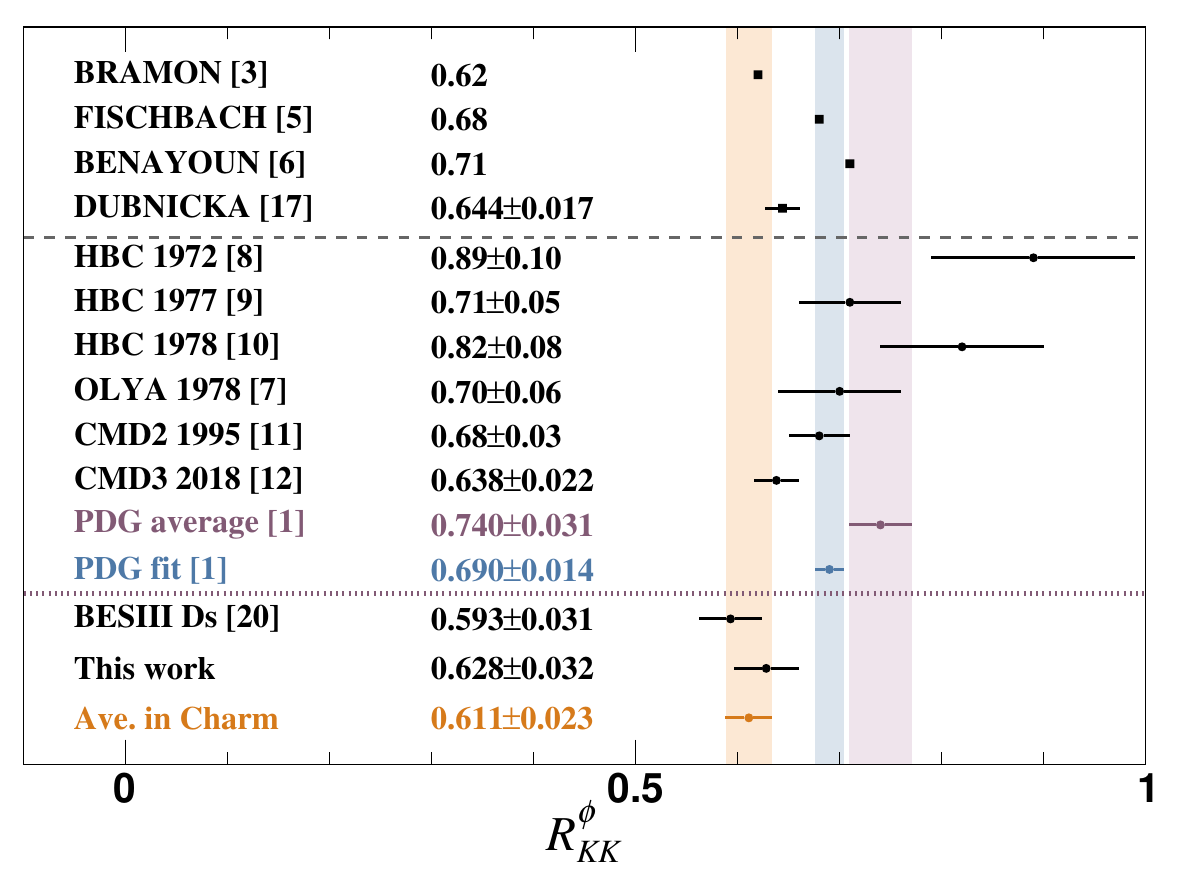}
    \caption{Comparison of the results for $R^{\phi}_{KK}$ measured in charmed
    meson decays, and the HBC, OLYA, CMD2, CMD3 experiments. Above the dashed 
    line are the theoretical calculations, below the experimental results. The
    yellow, purple and blue bands present the uncertainties of $R^{\phi}_{KK}$ 
    average measured  in charm mesons decays, world-average of measurements, and the PDG fit to all $\phi$ decay data, respectively.
    }
    \label{fig:comp}
  \end{center}
\end{figure}

Using the known value for 
${\cal B}(D^+ \to \phi \pi^+,\,\phi \to K^+K^-)=(2.69^{+0.07}_{-0.08})\times 10^{-3}$~\cite{PDG},
we find $R^{\phi}_{KK}=0.628 \pm 0.022 \pm 0.015 \pm 0.017$, where the third
uncertainty is that associated with
${\cal B}(D^+ \to \phi \pi^+,\,\phi \to K^+K^-)$. Combining this with our
previous measurement of $R^{\phi}_{KK}$ in $D_s^+$ decays~\cite{lihui}, we
obtain $0.611 \pm 0.023$ as the average of $R^{\phi}_{KK}$ measurements from
charm decays. Figure~\ref{fig:comp} compares our results with theoretical
predictions and measurements in $e^+e^-$ annihilation and $K-p$ scattering
experiments. The $R^{\phi}_{KK}$ average in charm is consistent with that
reported in Ref.~\cite{Dubn} but exhibits a significant tension of $3.4$
standard deviations with the current world-average of direct measurements.
The result disagrees by 2.5 standard deviations with the $R_{KK}^{\phi}$ result
obtained by the PDG from a fit to $\phi$ BF and partial width
measurements \cite{PDG}.
It should be noted that the current PDG average and
fit do not incorporate the CMD2 and CMD3 measurements, both of which are
individually consistent with the charm-based average reported here. In light of
this, a future comprehensive re-evaluation of all $\phi$ decay BFs by the PDG,
including the CMD2, CMD3 and BESIII results, would be most welcome. The 10 billion $J/\psi$ events collected by BESIII will also provide a sample of approximately $10^{6}\sim10^{7}$ $\phi$ mesons, enabling the measurement of $\phi$ BFs at the 0.5\% precision level.

\begin{figure}[t!]
  \begin{center}
    \includegraphics[width=0.6\textwidth]{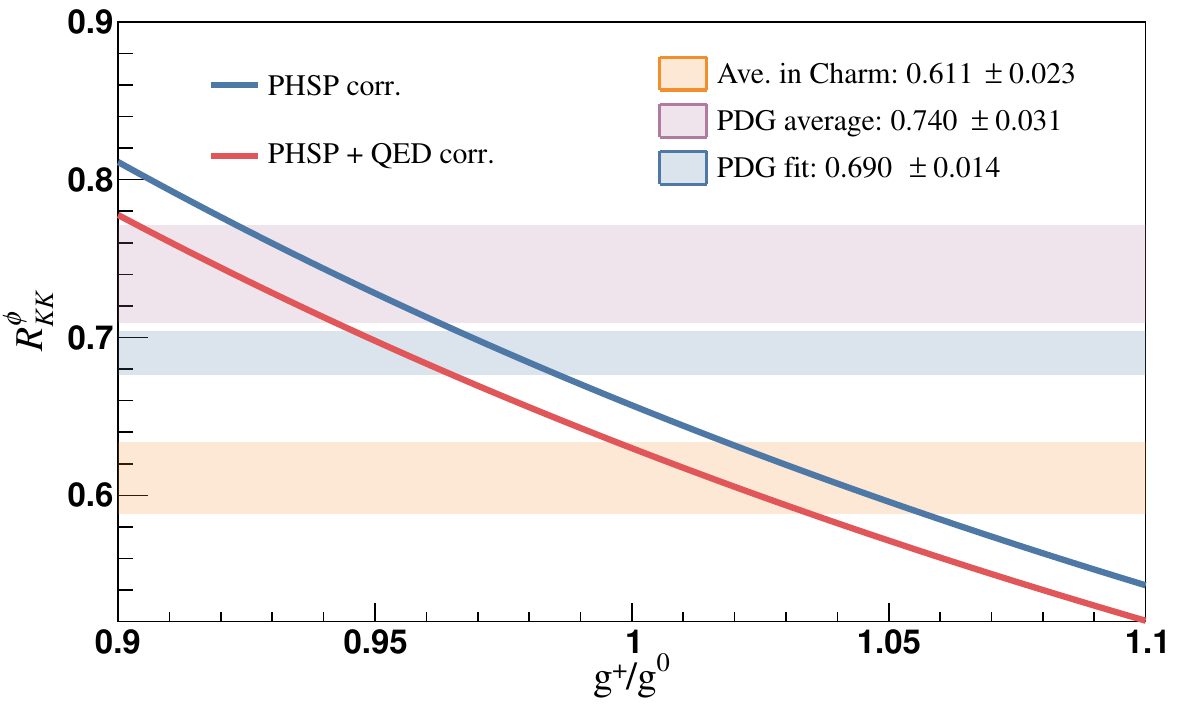}
    \caption{Comparison of measured $R^{\phi}_{KK}$ with different theoretical predictions~\cite{Flores, BRAMON} as a function of $g^+/g^0$. The differences between the theoretical curves are described in the text. 
    The colored bands show $\pm 1\sigma$ intervals.
 }
    \label{fig:comp_g}
  \end{center}
\end{figure}

Figure~\ref{fig:comp_g} shows $R^{\phi}_{KK}$ as a function of the 
 $\phi \to K^+K^-$ and $\phi \to K^0\bar{K}^0$ coupling ratio $g^+/g^0$, where the two curves are theoretical predictions~\cite{Flores, BRAMON} with corrections for phase-space difference (PHSP) and electromagnetic radiation (QED corr.). The PDG average and fit values imply a $g^+/g^0$ significantly less than one, which is difficult to reconcile with known isospin-breaking effects such as the $u$- and $d$-quark mass difference. 
However, our measurement of $R_{KK}^{\phi}$ from charm decay yields a $g^+/g^0$ consistent with unity. This concludes that PHSP and QED corrections are sufficient, thus rendering additional corrections unnecessary.
We also determine $\mathcal{B}(\phi \to K^+K^-)$  by combining the charm-based result for $R_{KK}^{\phi}$ with the average of our previous results for the ratio 
$\mathcal{B}(\phi\to\pi^+\pi^-\pi^0)/\mathcal{B}(\phi\to K^+K^-)=0.228 \pm 0.014$~\cite{xiaoyu, BESIII:2025nou} and the total BF of other known $\phi$ decays~\cite{PDG}. Using unitarity, we find $\mathcal{B}(\phi \to K^+K^-) = (53.54 \pm 0.76)\%$. This value is significantly higher than the values measured in $e^+e^-$ annihilation experiments~\cite{PDG}. Further measurements of the $\phi$ meson branching fractions are required to understand these differences.

Furthermore, we determine $\mathcal{B}(D^+ \to K_{S}^0K^{*}(892)^{+})$, $\mathcal{B}(D^{+} \to K_{L}^0K^{*}(892)^{+})$, and the phase-difference between the amplitudes for these decays.
The phase difference between the amplitudes for $D^{+} \to K_{S}^{0}K^{*}(892)^{+}$ and $D^{+} \to K_{L}^{0}K^{*}(892)^{+}$ is measured to be nearly $\pi$ radians, consistent with the theoretical expectation under $C\!P$ conservation.

\acknowledgments
The BESIII Collaboration thanks the staff of BEPCII (https://cstr.cn/31109.02.BEPC) and the IHEP computing center for their strong support. This work is supported in part by National Key R\&D Program of China under Contracts Nos. 2023YFA1606000, 2023YFA1606704, 2025YFA1613900; National Natural Science Foundation of China (NSFC) under Contracts Nos. 12605002, 12675104, 123B2077, 11635010, 11935015, 11935016, 11935018, 12025502, 12035009, 12035013, 12061131003, 12192260, 12192261, 12192262, 12192263, 12192264, 12192265, 12221005, 12225509, 12235017, 12342502, 12361141819; Joint Large-Scale Scientific Facility Fund of the NSFC and the Chinese Academy of Sciences under Contract No.~U2032104; the Chinese Academy of Sciences (CAS) Large-Scale Scientific Facility Program; the Strategic Priority Research Program of Chinese Academy of Sciences under Contract No. XDA0480600; CAS under Contract No. YSBR-101; the Excellent Youth Foundation of Henan Scientific Commitee under Contract No.~242300421044; the Postdoctoral
Fellowship Program and China Postdoctoral Science Foundation
under Grant No. BX2026035;
the Fundamental Research Funds for the Central Universities under Contract No. 010-63263110; 100 Talents Program of CAS; The Institute of Nuclear and Particle Physics (INPAC) and Shanghai Key Laboratory for Particle Physics and Cosmology; ERC under Contract No. 758462; German Research Foundation DFG under Contract No. FOR5327; Istituto Nazionale di Fisica Nucleare, Italy; Knut and Alice Wallenberg Foundation under Contracts Nos. 2021.0174, 2021.0299, 2023.0315; Ministry of Development of Turkey under Contract No. DPT2006K-120470; National Research Foundation of Korea under Contract No. NRF-2022R1A2C1092335; National Science and Technology fund of Mongolia; Polish National Science Centre under Contract No. 2024/53/B/ST2/00975; STFC (United Kingdom); Swedish Research Council under Contract No. 2019.04595; U. S. Department of Energy under Contract No. DE-FG02-05ER41374

\bibliographystyle{JHEP}
\bibliography{references}

@article{PDG,
  title = {Review of Particle Physics},
  author = {Navas, S. and Amsler, C. and Gutsche, T. and Hanhart, C. and Hern\'andez-Rey, J. J. and Louren\ifmmode \mbox{\c{c}}\else \c{c}\fi{}o, C. and Masoni, A. and Mikhasenko, M. and Mitchell, R. E. and Patrignani, C. and Schwanda, C. and Spanier, S. and Venanzoni, G. and Yuan, C. Z. and Agashe, K. and Aielli, G. and Allanach, B. C. and Alvarez-Mu\~niz, J. and Antonelli, M. and Aschenauer, E. C. and Asner, D. M. and Assamagan, K. and Baer, H. and Banerjee, Sw. and Barnett, R. M. and Baudis, L. and Bauer, C. W. and Beatty, J. J. and Beringer, J. and Bettini, A. and Biebel, O. and Black, K. M. and Blucher, E. and Bonventre, R. and Briere, R. A. and Buckley, A. and Burkert, V. D. and Bychkov, M. A. and Cahn, R. N. and Cao, Z. and Carena, M. and Casarosa, G. and Ceccucci, A. and Cerri, A. and Chivukula, R. S. and Cowan, G. and Cranmer, K. and Crede, V. and Cremonesi, O. and D'Ambrosio, G. and Damour, T. and de Florian, D. and de Gouv\^ea, A. and DeGrand, T. and Demers, S. and Demiragli, Z. and Dobrescu, B. A. and D'Onofrio, M. and Doser, M. and Dreiner, H. K. and Eerola, P. and Egede, U. and Eidelman, S. and El-Khadra, A. X. and Ellis, J. and Eno, S. C. and Erler, J. and Ezhela, V. V. and Fava, A. and Fetscher, W. and Fields, B. D. and Freitas, A. and Gallagher, H. and Gershon, T. and Gershtein, Y. and Gherghetta, T. and Gonzalez-Garcia, M. C. and Goodman, M. and Grab, C. and Gritsan, A. V. and Grojean, C. and Groom, D. E. and Gr\"unewald, M. and Gurtu, A. and Haber, H. E. and Hamel, M. and Hashimoto, S. and Hayato, Y. and Hebecker, A. and Heinemeyer, S. and Hikasa, K. and Hisano, J. and H\"ocker, A. and Holder, J. and Hsu, L. and Huston, J. and Hyodo, T. and Ianni, Al. and Kado, M. and Karliner, M. and Katz, U. F. and Kenzie, M. and Khoze, V. A. and Klein, S. R. and Krauss, F. and Kreps, M. and Kri\ifmmode \check{z}\else \v{z}\fi{}an, P. and Krusche, B. and Kwon, Y. and Lahav, O. and Lellouch, L. P. and Lesgourgues, J. and Liddle, A. R. and Ligeti, Z. and Lin, C.-J. and Lippmann, C. and Liss, T. M. and Lister, A. and Littenberg, L. and Lugovsky, K. S. and Lugovsky, S. B. and Lusiani, A. and Makida, Y. and Maltoni, F. and Manohar, A. V. and Marciano, W. J. and Matthews, J. and Mei\ss{}ner, U.-G. and Melzer-Pellmann, I.-A. and Mertsch, P. and Miller, D. J. and Milstead, D. and M\"onig, K. and Molaro, P. and Moortgat, F. and Moskovic, M. and Nagata, N. and Nakamura, K. and Narain, M. and Nason, P. and Nelles, A. and Neubert, M. and Nir, Y. and O'Connell, H. B. and O'Hare, C. A. J. and Olive, K. A. and Peacock, J. A. and Pianori, E. and Pich, A. and Piepke, A. and Pietropaolo, F. and Pomarol, A. and Pordes, S. and Profumo, S. and Quadt, A. and Rabbertz, K. and Rademacker, J. and Raffelt, G. and Ramsey-Musolf, M. and Richardson, P. and Ringwald, A. and Robinson, D. J. and Roesler, S. and Rolli, S. and Romaniouk, A. and Rosenberg, L. J and Rosner, J. L. and Rybka, G. and Ryskin, M. G. and Ryutin, R. A. and Safdi, B. and Sakai, Y. and Sarkar, S. and Sauli, F. and Schneider, O. and Sch\"onert, S. and Scholberg, K. and Schwartz, A. J. and Schwiening, J. and Scott, D. and Sefkow, F. and Seljak, U. and Sharma, V. and Sharpe, S. R. and Shiltsev, V. and Signorelli, G. and Silari, M. and Simon, F. and Sj\"ostrand, T. and Skands, P. and Skwarnicki, T. and Smoot, G. F. and Soffer, A. and Sozzi, M. S. and Spiering, C. and Stahl, A. and Sumino, Y. and Takahashi, F. and Tanabashi, M. and Tanaka, J. and Ta\ifmmode \check{s}\else \v{s}\fi{}evsk\'y, M. and Terao, K. and Terashi, K. and Terning, J. and Thoma, U. and Thorne, R. S. and Tiator, L. and Titov, M. and Tovey, D. R. and Trabelsi, K. and Urquijo, P. and Valencia, G. and Van de Water, R. and Varelas, N. and Verde, L. and Vivarelli, I. and Vogel, P. and Vogelsang, W. and Vorobyev, V. and Wakely, S. P. and Walkowiak, W. and Walter, C. W. and Wands, D. and Weinberg, D. H. and Weinberg, E. J. and Wermes, N. and White, M. and Wiencke, L. R. and Willocq, S. and Woody, C. L. and Workman, R. L. and Yao, W.-M. and Yokoyama, M. and Yoshida, R. and Zanderighi, G. and Zeller, G. P. and Zhu, R.-Y. and Zhu, S.-L. and Zimmermann, F. and Zyla, P. A. and Anderson, J. and Kramer, M. and Schaffner, P. and Zheng, W.},
  collaboration = {Particle Data Group Collaboration},
  journal = {Phys. Rev. D},
  volume = {110},
  issue = {3},
  pages = {030001},
  numpages = {5},
  year = {2024},
  month = {Aug},
  publisher = {American Physical Society},
  doi = {10.1103/PhysRevD.110.030001},
  url = {https://link.aps.org/doi/10.1103/PhysRevD.110.030001}
}

@article{Ablikim:2009aa,
title = {Design and construction of the {BESIII} detector},
journal = "Nucl. Instrum. Meth. A",
volume = {614},
number = {3},
pages = {345-399},
year = {2010},
issn = {0168-9002},
doi = {https://doi.org/10.1016/j.nima.2009.12.050},
url = {https://www.sciencedirect.com/science/article/pii/S0168900209023870},
author = {M. Ablikim and Z.H. An and J.Z. Bai and Niklaus Berger and J.M. Bian and X. Cai and G.F. Cao and X.X. Cao and J.F. Chang and C. Chen and G. Chen and H.C. Chen and H.X. Chen and J. Chen and J.C. Chen and L.P. Chen and P. Chen and X.H. Chen and Y.B. Chen and M.L. Chen and Y.P. Chu and X.Z. Cui and H.L. Dai and Z.Y. Deng and M.Y. Dong and S.X. Du and Z.Z. Du and J. Fang and C.D. Fu and C.S. Gao and M.Y. Gong and W.X. Gong and S.D. Gu and B.J. Guan and J. Guan and Y.N. Guo and J.F. Han and K.L. He and M. He and X. He and Y.K. Heng and Z.L. Hou and H.M. Hu and T. Hu and B. Huang and J. Huang and S.K. Huang and Y.P. Huang and Q. Ji and X.B. Ji and X.L. Ji and L.K. Jia and L.L. Jiang and X.S. Jiang and D.P. Jin and S. Jin and Y. Jin and Y.F. Lai and G.K. Lei and F. Li and G. Li and H.B. Li and H.S. Li and J. Li and J. Li and J.C. Li and Q.J. Li and L. Li and L. Li and R.B. Li and R.Y. Li and W.D. Li and W.G. Li and X.N. Li and X.P. Li and X.R. Li and Y.R. Li and W. Li and D.X. Lin and B.J. Liu and C.X. Liu and F. Liu and G.M. Liu and H. Liu and H.M. Liu and H.W. Liu and J.B. Liu and L.F. Liu and Q. Liu and Q.G. Liu and S.D. Liu and W.J. Liu and X. Liu and X.Z. Liu and Y. Liu and Y.J. Liu and Z.A. Liu and Z.Q. Liu and Z.X. Liu and J.G. Lu and T. Lu and Y.P. Lu and X.L. Luo and H.L. Ma and Q.M. Ma and X. Ma and X.Y. Ma and Z.P. Mao and J. Min and X.H. Mo and J. Nie and Z.D. Nie and R.G. Ping and S. Qian and Q. Qiao and G. Qin and Z.H. Qin and J.F. Qiu and R.G. Liu and Z.Y. Ren and G. Rong and L. Shang and D.L. Shen and X.Y. Shen and H.Y. Sheng and Y.F. Shi and L.W. Song and W.Y. Song and D.H. Sun and G.X. Sun and H.S. Sun and L.J. Sun and S.S. Sun and X.D. Sun and Y.Z. Sun and Z.J. Sun and J.P. Tan and S.Q. Tang and X. Tang and N. Tao and H.L. Tian and Y.R. Tian and X. Wan and D.Y. Wang and J.K. Wang and J.Z. Wang and K. Wang and K.X. Wang and L. Wang and L. Wang and L.J. Wang and L.S. Wang and M. Wang and N. Wang and P. Wang and P.L. Wang and Q. Wang and Y.F. Wang and Z. Wang and Z. Wang and Z.G. Wang and Z.Y. Wang and C.L. Wei and S.J. Wei and S.P. Wen and J.J. Wu and L.H. Wu and N. Wu and Y.H. Wu and Y.M. Wu and Z. Wu and M.H. Xu and X.M. Xia and H.S. Xiang and G. Xie and X.X. Xie and Y.G. Xie and G.F. Xu and H. Xu and Q.J. Xu and J.D. Xue and L. Xue and L. Yan and G.A. Yang and H. Yang and H.X. Yang and S.M. Yang and M. Ye and B.X. Yu and C. Yuan and C.Z. Yuan and Y. Yuan and S.L. Zang and B.X. Zhang and B.Y. Zhang and C.C. Zhang and C.C. Zhang and D.H. Zhang and H.Y. Zhang and J. Zhang and J.W. Zhang and J.Y. Zhang and L.S. Zhang and M. Zhang and Q.X. Zhang and W. Zhang and X.M. Zhang and Y. Zhang and Y.H. Zhang and Y.Y. Zhang and Z.X. Zhang and S.H. Zhang and D.X. Zhao and D.X. Zhao and H.S. Zhao and J.B. Zhao and J.W. Zhao and J.Z. Zhao and L. Zhao and P.P. Zhao and Y.B. Zhao and Y.D. Zhao and B. Zheng and J.P. Zheng and L.S. Zheng and Z.P. Zheng and B.Q. Zhou and G.M. Zhou and J. Zhou and L. Zhou and Z.L. Zhou and H.T. Zhu and K. Zhu and K.J. Zhu and Q.M. Zhu and X.W. Zhu and Y.S. Zhu and Z.A. Zhu and B.A. Zhuang and J.H. Zou and X. Zou and J.X. Zuo and L.L. Wang and M.H. Ye and Y.H. Zheng and C.F. Qiao and X.R. Lu and H.B. Liu and J.F. Hu and Y.T. Gu and X.D. Ruan and G.M. Huang and Y. Zeng and Y.H. Yan and G. Chelkov and I. Boyko and D. Dedovich and I. Denysenko and S. Grishin and A. Zhemchugov and Zhenjun Xiao and Jialun Ping and Libo {Guo Chenglin Luo} and Shenjian Chen and Ming Qi and Xiaowei Hu and Lei Zhang and Xueqian Li and Chunxu Yu and Yubin Liu and Ye Xu and Minggang Zhao and Aiqiang Guo and Yuping Guo and Zhenya He and Y.J. Mao and Z.Y. You and Y.T. Liang and X.Y. Zhang and X.T. Huang and J.B. Jiao and X.L. Li and M.Y. Duan and F.H. Liu and Q.W. Lu and F.P. Ning and X.D. Wang and Yongfei Liang and Changjian Tang and Yiyun Zhang and Y.N. Gao and H. Gong and B.B. Shao and Y.R. Tian and S.M. Yang and F.A. Harris and J.W. Kennedy and Q. Liu and X. Nguyen and S.L. Olsen and M. Rosen and C.P. Shen and G.S. Varner and X. Yu and Y. Zhou and H. Liang and Y. Chen and J. Xue and Q. Liu and B. Liu and Z. Cheng and L. Zhou and H. Yang and H.F. Chen and Cheng Li and M. Shao and Y.J. Sun and J. Yan and Z.B. Tang and X. Li and L. Zhao and L. Jiang and Z.P. Zhang and J. Wu and Z.Z. Xu and Q. Shan and Z. Xue and X.L. Wang and Q. An and S.B. Liu and J.H. Guo and L. Zhao and C.Q. Feng and X.Z. Liu and H. Li and W. Zheng and H. Yan and Z. Cao and X.H. Liu and Sachio Komamiya and Tomoyuki Sanuki and Taiki Yamamura and Tianchi Zhao and Mingxing Luo}
}

@article{Ablikim:2019hff,
    author = "Ablikim, M. and others",
    collaboration = "BESIII",
    title = "{Future physics programme of BESIII}",
    archivePrefix = "arXiv",
    primaryClass = "hep-ex",
    reportNumber = "HEP-Physics-Report-BESIII-2019-12-13",
    doi = "10.1088/1674-1137/44/4/040001",
    journal = "Chin. Phys. C",
    volume = "44",
    number = "4",
    pages = "040001",
    year = "2020"
}

@InProceedings{Yu:IPAC2016-TUYA01,
  author       = {C.H. Yu and others},
  title        = {{BEPCII} {P}erformance and {B}eam {D}ynamics {S}tudies on {L}uminosity},
  booktitle    = {Proc. of International Particle Accelerator Conference (IPAC'16),
                  Busan, Korea, May 8-13, 2016},
  pages        = {1014--1018},
  paper        = {TUYA01},
  language     = {english},
  venue        = {Busan, Korea},
  series       = {International Particle Accelerator Conference},
  number       = {7},
  publisher    = {JACoW},
  address      = {Geneva, Switzerland},
  month        = {June},
  year         = {2016},
  isbn         = {978-3-95450-147-2},
  doi          = {doi:10.18429/JACoW-IPAC2016-TUYA01},
  url          = {http://jacow.org/ipac2016/papers/tuya01.pdf},
  note         = {doi:10.18429/JACoW-IPAC2016-TUYA01},
}

@article{etof1,
    author = "Li, Xin and Sun, Yongjie and Li, Cheng and Liu, Zhen and Heng, Yuekun and Shao, Ming and Wang, Xiaozhuang and Wu, Zhi and Cao, Ping and Chen, Mingming and Dai, Hongliang and Liu, Shubing and Luo, Xiaolan and Jiang, Xiaoshan and Sun, Shengsen and Tang, Zebo and Sun, Weijia and Wang, Siyu and Xu, Meihang and Yang, Rongxing and Zhu, Kejun",
    title = "{Study of MRPC technology for BESIII endcap-TOF upgrade}",
    doi = "10.1007/s41605-017-0014-2",
    journal = "Radiat. Detect. Technol. Methods",
    volume = "1",
    pages = "13",
    year = "2017"
}

@article{etof2,
    author = "Guo, Ying-Xiao and Sun, Sheng-Sen and An, Fen-Fen and Yang, Rong-Xing and Zhou, Ming and Wu, Zhi and Dai, Hong-Liang and Heng, Yue-Kun and Li, Cheng and Deng, Zi-Yan and Liu, Huai-Min and Li, Wei-Guo",
    title = "{The study of time calibration for upgraded end cap TOF of BESIII}",
    doi = "10.1007/s41605-017-0012-4",
    journal = "Radiat. Detect. Technol. Methods",
    volume = "1",
    pages = "15",
    year = "2017"
}

@article{etof3,
title = {Design and construction of the new BESIII endcap Time-of-Flight system with MRPC Technology},
journal = {Nucl. Instrum. Meth. A},
volume = {953},
pages = {163053},
year = {2020},
issn = {0168-9002},
doi = {https://doi.org/10.1016/j.nima.2019.163053},
url = {https://www.sciencedirect.com/science/article/pii/S0168900219314068},
author = {P. Cao and H.F. Chen and M.M. Chen and H.L. Dai and Y.K. Heng and X.L. Ji and X.S. Jiang and C. Li and X. Li and S.B. Liu and Z. Liu and X.L. Luo and X.Y. Ma and M. Shao and S.S. Sun and Y.J. Sun and Z.B. Tang and X.Z. Wang and Z. Wu and M.H. Xu and R.X. Yang and M. Ye and J. Zhang and Y.H. Zhang and J.Z. Zhao}
}

@article{GEANT4,
    author = "Agostinelli, S. and others",
    collaboration = "GEANT4",
    title = "{GEANT4 -- a simulation toolkit}",
    reportNumber = "SLAC-PUB-9350, FERMILAB-PUB-03-339, CERN-IT-2002-003",
    doi = "10.1016/S0168-9002(03)01368-8",
    journal = "Nucl. Instrum. Meth. A",
    volume = "506",
    pages = "250--303",
    year = "2003"
}

@article{KKMC1,
  title = {Coherent exclusive exponentiation for precision {M}onte {C}arlo calculations},
  author = {Jadach, S. and Ward, B. F. L. and Wa\ifmmode \mbox{\c{}}\else \c{}\fi{}s, Z.},
  journal = {Phys. Rev. D},
  volume = {63},
  issue = {11},
  pages = {113009},
  numpages = {65},
  year = {2001},
  month = {May},
  publisher = {American Physical Society},
  doi = {10.1103/PhysRevD.63.113009},
  url = {https://link.aps.org/doi/10.1103/PhysRevD.63.113009}
}

@article{KKMC2,
title = {The precision Monte Carlo event generator {KK} for two-fermion final states in $e^+e^-$ collisions},
journal = {Computer Physics Communications},
volume = {130},
number = {3},
pages = {260-325},
year = {2000},
issn = {0010-4655},
doi = {https://doi.org/10.1016/S0010-4655(00)00048-5},
url = {https://www.sciencedirect.com/science/article/pii/S0010465500000485},
author = {Jadach, S. and Ward, B. F. L. and Wa\ifmmode \mbox{\c{}}\else \c{}\fi{}s, Z.}
}

@article{EVTGEN1,
    author = "Lange, D.J.",
    editor = "Erhan, S. and Schlein, P. and Rozen, Y.",
    title = "{The EvtGen particle decay simulation package}",
    doi = "10.1016/S0168-9002(01)00089-4",
    journal = "Nucl. Instrum. Meth. A",
    volume = "462",
    pages = "152--155",
    year = "2001"
}

@article{EVTGEN2,
        doi = {10.1088/1674-1137/32/8/001},
        url = {https://doi.org/10.1088%2F1674-1137%2F32%2F8%2F001},
        year = 2008,
        month = {aug},
        publisher = {{IOP} Publishing},
        volume = {32},
        number = {8},
        pages = {599--602},
        author = {Ping, Rong-Gang},
        title = {Event generators at {BESIII}},
        journal = {Chin. Phys. C}
}

@article{LUNDCHARM1,
  title = "{Event generator for $J/\ensuremath{\psi}$ and $\ensuremath{\psi}(2S)$ decay}",
  author = {Chen, J. C. and Huang, G. S. and Qi, X. R. and Zhang, D. H. and Zhu, Y. S.},
  journal = {Phys. Rev. D},
  volume = {62},
  issue = {3},
  pages = {034003},
  numpages = {8},
  year = {2000},
  month = {Jun},
  publisher = {American Physical Society},
  doi = {10.1103/PhysRevD.62.034003},
  url = {https://link.aps.org/doi/10.1103/PhysRevD.62.034003}
}

@article{LUNDCHARM2,
author = {Yang, Rui-Ling and Ping, Rong-Gang and Chen, Hong},
title = "{Tuning and validation of the Lundcharm model with $J/\psi$ decays}",
publisher = {Chin. Phys. Lett.},
year = {2014},
journal = {Chin. Phys. Lett.},
volume = {31},
number = {06},
eid = {061301},
numpages = {0},
pages = {061301},
url = {http://cpl.iphy.ac.cn/EN/abstract/article_60019.shtml},
doi = {10.1088/0256-307X/31/6/061301}
}

@article{PHOTOS,
title = "{QED bremsstrahlung in semileptonic B and leptonic $\tau$ decays}",
journal = "Phys. Lett. B",
volume = "303",
number = "1",
pages = "163 - 169",
year = "1993",
issn = "0370-2693",
doi = "https://doi.org/10.1016/0370-2693(93)90062-M",
url = "http://www.sciencedirect.com/science/article/pii/037026939390062M",
author = "E. Richter-Was"
}

@article{EventFilter,
    author = "Zhang, Jia-Wei and others",
    title = "{Suppression of top-up injection backgrounds with offline event filter in the BESIII experiment}",
    doi = "10.1007/s41605-022-00331-7",
    journal = "Radiat. Detect. Technol. Methods",
    volume = "6",
    number = "3",
    pages = "289--293",
    year = "2022"
}

@article{lum2,
    author = "Ablikim, Medina and others",
    collaboration = "BESIII",
    title = "{Measurement of integrated luminosity of data collected at 3.773 GeV by BESIII from 2021 to 2024}",
    archivePrefix = "arXiv",
    primaryClass = "hep-ex",
    doi = "10.1088/1674-1137/ad70a0",
    journal = "Chin. Phys. C",
    volume = "48",
    number = "12",
    pages = "123001",
    year = "2024"
}

@article{EcmsMea,
title={Online monitoring of the center-of-mass energy from real data at {BESIII}},
author={Lu, Jiada and Xiao, Yanjia and Ji, Xiaobin},
journal={Radiation Detection Technology and Methods},
volume={4},
number={3},
pages={337-344},
year={2020},
doi={10.1007/s41605-020-00188-8},
issn={2509-9949},
url={https://doi.org/10.1007/s41605-020-00188-8}
}

@article{DTmethod,
    author = "Baltrusaitis, R. M. and others",
    collaboration = "MARK-III",
    title = "{Direct Measurements of Charmed-$D$-Meson Hadronic Branching Fractions}",
    reportNumber = "SLAC-PUB-3861",
    doi = "10.1103/PhysRevLett.56.2140",
    journal = "Phys. Rev. Lett.",
    volume = "56",
    pages = "2140",
    year = "1986"
}

@article{cremmer,
  title   = "{Radiative corrections to $\phi$ decay}",
  author  = {Cremmer, E. and Gourdin, M.},
  journal = {Nucl. Phys. B},
  volume  = {9},
  pages   = {451--459},
  year    = {1969},
  doi     = {10.1016/0550-3213(69)90141-2}
}

@article{BRAMON,
  title   = "{The ratio $\Gamma(\phi\to K^+K^-)/\Gamma(\phi\to K^0\bar{K}^0)$}",
  author  = {Bramon, A. and Escribano, R. and Lucio M, J. L. and Pancheri, G.},
  journal = {Phys. Lett. B},
  volume  = {486},
  pages   = {406--413},
  year    = {2000},
  doi     = {10.1016/S0370-2693(00)00770-X}
}

@article{Flores,
  title   = "{Structure-dependent radiative corrections to $\phi\to K^+K^-/K_LK_S$ decays}",
  author  = {Flores-Ba\'ez, F. V. and L\'opez Castro, G.},
  journal = {Phys. Rev. D},
  volume  = {78},
  pages   = {077301},
  year    = {2008},
  doi     = {10.1103/PhysRevD.78.077301}
}

@article{Fischbach,
  title   = "{Corrections to Fermi's golden rule in $\phi\to K\bar{K}$ decays}",
  author  = {Fischbach, E. and Overhauser, A. W. and Woodahl, B.},
  journal = {Phys. Lett. B},
  volume  = {526},
  pages   = {355},
  year    = {2002},
  doi     = {10.1016/S0370-2693(01)01520-9}
}

@article{Benayoun,
  title   = "{Upgraded breaking of the HLS model: a full solution to the $\tau-e^+e^-$ and $\phi$ decay issues and its consequences on $g-2$ VMD estimates}",
  author  = {Benayoun, M. and David, P. and DelBuono, L. and Jegerlehner, F.},
  journal = {Eur. Phys. J. C},
  volume  = {72},
  pages   = {1848},
  year    = {2012},
  doi     = {10.1140/epjc/s10052-011-1848-2}
}

@article{OLYA,
  title   = "{The $\phi$ meson: Precision measurement of its mass, observation of the $\omega$-$\phi$ interference}",
  author  = {Bukin, A. D. and Kurdadze, L. M. and Serednyakov, S. I. and Sidorov, V. A. and Skrinsky, A. N. and Shatunov, Yu. M. and Shvarts, B. A. and Eidelman, S. I.},
  journal = {Sov. J. Nucl. Phys.},
  volume  = {27},
  pages   = {516},
  year    = {1978},
  url     = {https://inspirehep.net/literature/137035}
}

@article{HBC72,
  title   = "{Study of nonstrange mesons produced in $K^-p$ interactions at 3.9 and 4.6 GeV/$c$}",
  author  = {Benitez, M. A. and Chung, S. U. and Eisner, R. L. and Samios, N. P.},
  journal = {Phys. Rev. D},
  volume  = {6},
  pages   = {29},
  year    = {1972},
  doi     = {10.1103/PhysRevD.6.29}
}

@article{HBC77,
  title   = "{Study of the reactions $K^-p \to K^+K^-\Lambda$ and $K^-p \to pp\Lambda$ at 10 and 16 GeV/$c$}",
  author  = {Laven, H. and others},
  collaboration = {Aachen-Berlin-CERN-London-Vienna},
  journal = {Nucl. Phys. B},
  volume  = {127},
  pages   = {43},
  year    = {1977},
  doi     = {10.1016/0550-3213(77)90350-9}
}

@article{HBC78,
  title   = "{Non-strange vector meson production in $K^-p$ interactions at 4.2 GeV/$c$}",
  author  = {Losty, M. J. and others},
  collaboration = {Amsterdam-CERN-Nijmegen-Oxford},
  journal = {Nucl. Phys. B},
  volume  = {133},
  pages   = {38},
  year    = {1978},
  doi     = {10.1016/0550-3213(78)90168-2}
}

@article{CMD2,
  title   = "{Measurement of $\phi$ meson parameters with CMD-2 detector at VEPP-2M collider}",
  author  = {Akhmetshin, R. R. and others},
  journal = {Phys. Lett. B},
  volume  = {364},
  pages   = {199},
  year    = {1995},
  doi     = {10.1016/0370-2693(95)01394-6}
}

@article{CMD3,
  title   = "{Study of the process $e^+e^-\to K^+K^-$ in the center-of-mass energy range 1010--1060 MeV with the CMD-3 detector}",
  author  = {Kozyrev, E. A. and others},
  journal = {Phys. Lett. B},
  volume  = {779},
  pages   = {64},
  year    = {2018},
  doi     = {10.1016/j.physletb.2018.01.079}
}

@article{Parrour,
  title   = "{Evidence for an interference effect between $\omega$ and $\phi$ resonances in $\pi^+\pi^-\pi^0$ production by $e^+e^-$ annihilation}",
  author  = {Parrour, G. and Cosme, G. and Courau, A. and Dudelzak, B. and others},
  journal = {Phys. Lett. B},
  volume  = {63},
  pages   = {357},
  year    = {1976},
  doi     = {10.1016/0370-2693(76)90282-3}
}

@article{Mattiuzzi,
  title   = "{Thermal fluctuation and collisional damping effects in the GDR observables in hot rotating $^{176}$W nuclei}",
  author  = {Mattiuzzi, M. and Bracco, A. and Camera, F. and Million, B. and others},
  journal = {Phys. Lett. B},
  volume  = {364},
  pages   = {13},
  year    = {1995},
  doi     = {10.1016/0370-2693(95)01230-5}
}

@article{Dolinsky,
  title   = "{Summary of experiments with the neutral detector at the $e^+e^-$ storage ring VEPP-2M}",
  author  = {Dolinsky, S. I. and Druzhinin, V. P. and Dubrovin, M. S. and Golubev, V. B. and others},
  journal = {Phys. Rept.},
  volume  = {202},
  pages   = {99},
  year    = {1991},
  doi     = {10.1016/0370-1573(91)90127-8}
}

@article{Dubn,
  title   = "{What is the real value of the ratio $R=\phi\to K^+K^- / K_L^0K_S^0$?}",
  author  = {Dubni{\v c}ka, S. and Dubni{\v c}kov{\'a}, A. Z. and Holka, L. and Liptaj, A.},
  journal = {Phys. Rev. D},
  volume  = {110},
  pages   = {054019},
  year    = {2024},
  doi     = {10.1103/PhysRevD.110.054019}
}

@article{xiaoyu,
  title   = "{Study of $D_s^+\to f_0(980)\rho^+$ and $\phi\pi^+$ decays through $D_s^+\to\pi^+\pi^+\pi^-\pi^0$}",
  author  = {Ablikim, M. and others},
  collaboration = {BESIII},
  journal = {Phys. Rev. Lett.},
  volume  = {134},
  pages   = {011904},
  year    = {2025},
  doi     = {10.1103/PhysRevLett.134.011904}
}

@article{BESIII:2025nou,
  title   = "{Observation of the $W$-annihilation process $D_s^+\to\omega\rho^+$ and measurement of $D_s^+\to\phi\rho^+$ in $D_s^+\to\pi^+\pi^+\pi^-\pi^0\pi^0$ decays}",
  author  = {Ablikim, M. and others},
  collaboration = {BESIII},
  journal = {Phys. Rev. Lett.},
  volume  = {134},
  pages   = {201902},
  year    = {2025},
  doi     = {10.1103/PhysRevLett.134.201902}
}

@article{lihui,
  title   = "{Study of $\phi\to K\bar{K}$ and $K_S^0$--$K_L^0$ asymmetry in the amplitude analysis of $D_s^+\to K_S^0K_L^0\pi^+$ decays}",
  author  = {Ablikim, M. and others},
  collaboration = {BESIII},
  journal = {Phys. Rev. Lett.},
  volume  = {135},
  pages   = {161902},
  year    = {2025},
  doi     = {10.1103/6py9-h8qv}
}

@article{Wang,
  title   = "{$K_S^0$--$K_L^0$ asymmetries in $D$-meson decays}",
  author  = {Wang, D. and Yu, F. S. and Guo, P. F. and Jiang, H. Y.},
  journal = {Phys. Rev. D},
  volume  = {95},
  pages   = {073007},
  year    = {2017},
  doi     = {10.1103/PhysRevD.95.073007}
}

@article{cheng,
  title   = "{Updated analysis of $D\to PP$, $VP$, and $VV$ decays: Implications for $K_S^0$--$K_L^0$ asymmetries and $D^0$-$\bar{D}^0$ mixing}",
  author  = {Cheng, Hai-Yang and Chiang, Cheng-Wei},
  journal = {Phys. Rev. D},
  volume  = {109},
  pages   = {073008},
  year    = {2024},
  doi     = {10.1103/PhysRevD.109.073008}
}

@article{Qin:2013tje,
  title   = "{Branching ratios and direct $CP$ asymmetries in $D\to PV$ decays}",
  author  = {Qin, Q. and Li, H. N. and Lyu, C. D. and Yu, F. S.},
  journal = {Phys. Rev. D},
  volume  = {89},
  pages   = {054006},
  year    = {2014},
  doi     = {10.1103/PhysRevD.89.054006}
}

@article{kskpi0,
  title   = "{Measurements of the absolute branching fractions and $CP$ asymmetries for $D^+\to K_{S,L}^0K^+(\pi^0)$}",
  author  = {Ablikim, M. and others},
  collaboration = {BESIII},
  journal = {Phys. Rev. D},
  volume  = {99},
  pages   = {032002},
  year    = {2019},
  doi     = {10.1103/PhysRevD.99.032002}
}

@article{BESIII:2024ncc,
  title   = "{Study of $D^+\to K_S^0K^*(892)^+$ in $D^+\to K_S^0K_S^0\pi^+$}",
  author  = {Ablikim, M. and others},
  collaboration = {BESIII},
  journal = {Phys. Rev. D},
  volume  = {110},
  pages   = {092006},
  year    = {2024},
  doi     = {10.1103/PhysRevD.110.092006}
}

@article{ref:kspieta,
  title   = "{Observation of $D^+\to K_S^0a_0(980)^+$ in the amplitude analysis of $D^+\to K_S^0\pi^+\eta$}",
  author  = {Ablikim, M. and others},
  collaboration = {BESIII},
  journal = {Phys. Rev. Lett.},
  volume  = {132},
  pages   = {131903},
  year    = {2024},
  doi     = {10.1103/PhysRevLett.132.131903}
}

@article{BESIII:2019ymv,
  title   = "{Amplitude analysis of $D^+\to K_S^0\pi^+\pi^+\pi^-$}",
  author  = {Ablikim, M. and others},
  collaboration = {BESIII},
  journal = {Phys. Rev. D},
  volume  = {100},
  pages   = {072008},
  year    = {2019},
  doi     = {10.1103/PhysRevD.100.072008}
}

@article{covariant-tensors,
  title   = "{Covariant tensor formalism for partial-wave analyses of $\psi$ decay to mesons}",
  author  = {Zou, B. S. and Bugg, D. V.},
  journal = {Eur. Phys. J. A},
  volume  = {16},
  pages   = {537--547},
  year    = {2003},
  doi     = {10.1140/epja/i2002-10135-4}
}

@article{RBW,
  title   = "{Remarks on the phenomenological analysis of resonances}",
  author  = {Jackson, J. D.},
  journal = {Nuovo Cimento},
  volume  = {34},
  pages   = {1644},
  year    = {1964}
}

@book{Blatt,
  title     = "{Theoretical Nuclear Physics}",
  author    = {Blatt, J. M. and Weisskopf, V. F.},
  publisher = {John Wiley \& Sons},
  address   = {New York},
  year      = {1973}
}

@article{xgboost1,
  title   = "{Reweighting with boosted decision trees}",
  author  = {Rogozhnikov, A.},
  journal = {J. Phys. Conf. Ser.},
  volume  = {762},
  pages   = {012036},
  year    = {2016},
  doi     = {10.1088/1742-6596/762/1/012036}
}

@article{xgboost2,
  title   = "{Application of XGBoost to particle identification at BESIII}",
  author  = {Liu, B. and Xiong, X. and Hou, G. and Song, S. and Shen, L.},
  journal = {EPJ Web Conf.},
  volume  = {214},
  pages   = {06033},
  year    = {2019}
}

@article{BESIII:2025sea,
  title   = "{Amplitude analysis and branching fraction measurement of the decay $D^0\to K_S^0\pi^0\pi^0$}",
  author  = {Ablikim, M. and others},
  collaboration = {BESIII},
  journal = {arXiv},
  eprint  = {2510.25111},
  archivePrefix = {arXiv},
  primaryClass  = {hep-ex},
  year    = {2025}
}

@article{ref:Kspipi0,
  title   = "{Amplitude analysis and branching-fraction measurement of $D_s^+\to K_S^0\pi^+\pi^0$}",
  author  = {Ablikim, M. and others},
  collaboration = {BESIII},
  journal = {J. High Energ. Phys.},
  volume  = {06},
  pages   = {181},
  year    = {2021},
  doi     = {10.1007/JHEP06(2021)181}
}

@article{ref:KsKpipi,
  title   = "{Amplitude analysis and branching-fraction measurement of $D_s^+\to K_S^0K^-\pi^+\pi^+$}",
  author  = {Ablikim, M. and others},
  collaboration = {BESIII},
  journal = {Phys. Rev. D},
  volume  = {103},
  pages   = {092006},
  year    = {2021},
  doi     = {10.1103/PhysRevD.103.092006}
}

@article{ref:KsKpi0,
  title   = "{Observation of an $a_0$-like state with mass of 1.817 GeV in the study of $D_s^+\to K_S^0K^+\pi^0$ decays}",
  author  = {Ablikim, M. and others},
  collaboration = {BESIII},
  journal = {Phys. Rev. Lett.},
  volume  = {129},
  pages   = {182001},
  year    = {2022},
  doi     = {10.1103/PhysRevLett.129.182001}
}

@article{KpiLASS,
  title   = "{Measurement of $\cos 2\beta$ in $B^0\to D^{(*)}h^0$ with $D\to K_S^0\pi^+\pi^-$ decays by a combined time-dependent Dalitz plot analysis of $BABAR$ and Belle data}",
  author  = {Link, J. M. and others},
  collaboration = {FOCUS},
  journal = {Phys. Rev. D},
  volume  = {98},
  pages   = {112012},
  year    = {2018},
  doi   = {10.1103/PhysRevD.98.112012}
}

@article{Kmatrix,
  title   = "{K-matrix analysis of the $K\pi$ S-wave in the mass region 900--2100 MeV and nonet classification of scalar $q\bar{q}$ states}",
  author  = {Anisovich, A. V. and Sarantsev, A. V.},
  journal = {Phys. Lett. B},
  volume  = {413},
  pages   = {137--146},
  year    = {1997},
  doi     = {10.1016/S0370-2693(97)01089-7}
}

@article{BES:2007cev,
  title   = "{Direct measurements of the non-$D\bar{D}$ cross section $\sigma_{\psi(3770)\to non-D\bar{D}}$ at $E_{cm}=3.773$ GeV and the branching fraction for $\psi(3770)\to non-D\bar{D}$}",
  author  = {Ablikim, M. and others},
  collaboration = {BESIII},
  journal = {Phys. Rev. D},
  volume  = {76},
  pages   = {122002},
  year    = {2007},
  doi     = {10.1103/PhysRevD.76.122002}
}

@article{ke,
  title   = "{Amplitude analysis of charmed-meson decays at BESIII}",
  author  = {Zhang, H. and Li, Z. X. and Guan, C. Y. and Lu, Z. H. and others},
  journal = {Chin. Phys. C},
  volume  = {50},
  pages   = {093002},
  year    = {2025},
  doi     = {10.1088/1674-1137/ae7962}
}

@article{xxh,
  title   = "{Study of the decay $D^+\to K^*(892)^+K_S^0$ in $D^+\to K^+K_S^0\pi^0$}",
  author  = {Ablikim, M. and others},
  collaboration = {BESIII},
  journal = {Phys. Rev. D},
  volume  = {104},
  pages   = {012006},
  year    = {2021},
  doi     = {10.1103/PhysRevD.104.012006}
}

\clearpage
\appendix
\newcommand{\BESIIIorcid}[1]{\href{https://orcid.org/#1}{\hspace*{0.1em}\raisebox{-0.45ex}{\includegraphics[width=1em]{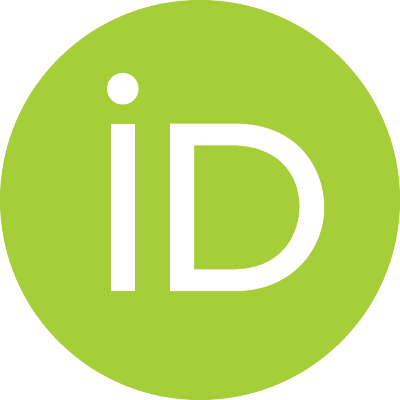}}}}
M.~Ablikim$^{1}$\BESIIIorcid{0000-0002-3935-619X},
M.~N.~Achasov$^{4,c}$\BESIIIorcid{0000-0002-9400-8622},
P.~Adlarson$^{83}$\BESIIIorcid{0000-0001-6280-3851},
X.~C.~Ai$^{89}$\BESIIIorcid{0000-0003-3856-2415},
C.~S.~Akondi$^{31A,31B}$\BESIIIorcid{0000-0001-6303-5217},
R.~Aliberti$^{39}$\BESIIIorcid{0000-0003-3500-4012},
A.~Amoroso$^{82A,82C}$\BESIIIorcid{0000-0002-3095-8610},
Q.~An$^{79,65,\dagger}$,
Y.~H.~An$^{89}$\BESIIIorcid{0009-0008-3419-0849},
Y.~Bai$^{63}$\BESIIIorcid{0000-0001-6593-5665}
% ...
\\
\vspace{0.2cm}
(BESIII Collaboration)\\
\vspace{0.2cm} {\it
$^{1}$ Institute of High Energy Physics, Beijing 100049, People's Republic of China\\
% ...
$^{4}$ Budker Institute of Nuclear Physics SB RAS (BINP), Novosibirsk 630090, Russia\\
% ...
$^{31}$ INFN Sezione di Ferrara, (A)INFN Sezione di Ferrara, I-44122, Ferrara, Italy; (B)University of Ferrara, I-44122, Ferrara, Italy\\
% ...
$^{39}$ Johannes Gutenberg University of Mainz, Johann-Joachim-Becher-Weg 45, D-55099 Mainz, Germany\\
% ...
$^{63}$ Southeast University, Nanjing 211100, People's Republic of China\\
% ...
$^{65}$ State Key Laboratory of Particle Detection and Electronics, Beijing 100049, Hefei 230026, People's Republic of China\\
% ...
$^{79}$ University of Science and Technology of China, Hefei 230026, People's Republic of China\\
% ...
$^{82}$ University of Turin and INFN, (A)University of Turin, I-10125, Turin, Italy; (B)University of Eastern Piedmont, I-15121, Alessandria, Italy; (C)INFN, I-10125, Turin, Italy\\
$^{83}$ Uppsala University, Box 516, SE-75120 Uppsala, Sweden\\
% ...
$^{89}$ Zhengzhou University, Zhengzhou 450001, People's Republic of China\\

\vspace{0.2cm}
$^{\dagger}$ Deceased\\
% ...
$^{c}$ Also at the Novosibirsk State University, Novosibirsk, 630090, Russia\\
% ...

}

\end{document}